\documentclass[11pt]{amsart}

\usepackage{amsmath,amssymb}
\usepackage{a4wide}
\usepackage{bbm}
\usepackage{graphics,graphicx,epsfig,pstcol,psfrag}  
\usepackage[hang,scriptsize,nooneline]{subfigure} 
\usepackage[T1]{fontenc}
\usepackage[latin1]{inputenc}

\newtheorem{theorem}{Theorem}
\newtheorem{prop}{Proposition}
\newtheorem{lemma}{Lemma}
\newtheorem{coro}{Corollary}
\newtheorem{fact}{Fact}

\theoremstyle{definition}
\newtheorem{remark}{Remark}

\def\EE{\mathbb{E}}
\def\PP{\mathbb{P}}

\def\RR{\mathbb{R}}
\def\ZZ{\mathbb{Z}}

\def\hZ{\widehat{Z}}

\newcommand{\btimes}{\mbox{\huge \raisebox{-0.2ex}{$\times$}}} 
\newcommand{\one}{\mathbbm 1}

\def\be{\begin{equation} \label}
\def\ee{\end{equation}}

\def\cM{\mathcal{M}}

\begin{document}
\bibliographystyle{abbrv}








\title[Single-crossover dynamics]
{Single-crossover dynamics: \\ finite versus infinite populations}

\author{Ellen  Baake and Inke Hildebrandt}
\address{Faculty of Technology,  Bielefeld University,
33594 Bielefeld, Germany}
\email{\{ebaake,ihildebr\}@techfak.uni-bielefeld.de}

\begin{abstract}
Populations  evolving under the joint influence 
of recombination and resampling (traditionally known as genetic drift)
are investigated. 
First, we summarise and adapt a deterministic approach, as valid for infinite 
populations, which assumes  continuous time and 
single crossover events. The corresponding nonlinear system of 
differential equations permits a closed solution, both in terms of the
type frequencies and via linkage disequilibria
of all orders. To include stochastic effects, we
then consider the corresponding finite-population model, 
the Moran model with single crossovers, and examine it both analytically
and by means of simulations. Particular emphasis is on the connection
with the deterministic solution.
If there is only recombination and every pair of recombined offspring
replaces their pair of parents
(i.e., there is no re\-sampling), 
then the {\em expected} type frequencies in the finite population,
of arbitrary size,  equal  the type frequencies in the infinite population.
If resampling is included, the stochastic process
converges, in the infinite-population limit,
to the deterministic dynamics, 
which turns out to be a good approximation already for 
populations of moderate size.
\end{abstract}

\maketitle





\section{Introduction}

It is well known that the dynamics of populations under recombination is 
notoriously difficult to treat, even if the population is so large that 
stochastic 
fluctuations due to genetic drift can be neglected, so that the time evolution 
may be described by a deterministic dynamical system. The difficulty lies 
in the inherent nonlinearity of the corresponding (difference or
differential) equations, which stems from the interaction of pairs of
parental individuals during recombination.

The overwhelming part of the literature on the dynamics of recombination 
deals with \emph{discrete} time. The first  solutions go back 
to Geiringer (1944, \cite{Geir44}) and Bennett (1954, \cite{Benn54}); their
construction was more recently worked out in greater detail and
completeness by Dawson \cite{Daws02}.
The underlying mathematical structures were investigated
within the framework of genetic algebras, see \cite{Lyub92,McHR83},
or \cite{Ring85}. Quite generally, the 
solution relies on a certain nonlinear transformation (known as Haldane
linearisation) from (gamete or haplotype) frequencies to suitable linkage disequilibria, which 
decouple from each other and decay geometrically. But if more than three loci 
are involved, this transformation must be constructed via recursions that 
involve the parameters of the recombination process, and is  not 
available explicitly except in the trivial case of free recombination
(i.e., independent gene loci). 
A different approach was taken by Barton and Turelli \cite{BaTu91} 
by translating type frequencies into certain sets of moments; the 
resulting iterations are 
well-suited for symbolic manipulation,  numerical treatment,
and biological interpretation, but do not 
lead to (and do not primarily aim at) explicit solutions. For a 
review of the area, see \cite[Ch.~V.4]{Buer00}.

It has  turned out recently \cite{Baak01,BaBa03,Baak05} that, 
in contrast to the 
discrete-time situation, the corresponding dynamics in \emph{continuous} time 
permits a simple explicit solution in a biologically relevant 
special case, namely, the situation in which at most one crossover
happens at any given time. Here,
only recombination events  may occur that partition the 
sites of a sequence (or the loci on a chromosome) into two {\em ordered} 
subsets 
that correspond to the sites before and after a given crossover point.
In contrast to previous approaches, the solution is given directly in terms
of the original type frequencies; but, again,
a transformation to certain linkage disequilibria is available
that linearise and diagonalise the system. The main 
simplification lies in the fact that
the transformation is independent 
of the recombination rates and is available in a simple explicit form.

However, the focus in modern population genetics is on {\em finite}
populations, 
which also experience genetic drift (that is, resampling).
In order to investigate the relevance
of the deterministic solution for the more involved stochastic system,
we shall, in this paper, examine the corresponding finite population model,
namely, the Moran model with recombination, and compare some of its
properties with the infinite population model, both analytically and
by means of simulations. We shall show the following:
\begin{itemize}
\item
In a {\em finite} population with recombination, but without resampling
(i.e.,  every pair of recombined offspring replaces its pair of
parents),
the type frequencies are,  {\em in expectation}, 
given by the corresponding quantities for {\em infinite} populations.
This property holds exactly as long as the genetic material is 
conserved in the recombination process, and to a very good
approximation otherwise. 
This connection between stochastic and deterministic models
is by no means obvious: It is usually reserved for
populations of individuals that evolve independently 
(like branching processes); or to
systems with interactions that
do not change the expected type frequencies (such as the Moran model 
with independent mutation and resampling). 
In contrast, interaction due to recombination {\em does} change
the expected composition of the population, and does, therefore, not 
appear to be of this simple form.

\item
If the joint action of recombination and resampling is considered, the
process of type frequencies differs from the deterministic dynamics,
even in expectation, particularly if the population  is small.
It follows from  general arguments, however, that the former 
converges to the latter when population size goes to infinity. By means
of simulations, we show that this limit yields a good
approximation of realistic biological
situations:
Convergence is fast enough to justify the use of the deterministic 
solution already for populations of moderate  size
(of the order of $10^5$ individuals, say).
\end{itemize}

The paper is organised as follows. We first recapitulate the 
single-crossover differential equation and its solution. This is 
appropriate 
since the original articles \cite{Baak05,BaBa03} use a rather abstract  
measure-theoretical framework suitable for very general type spaces,
which is not easily accessible to theoretical biologists. 
We  reduce the formalism  to the  important special
case of  finite type spaces, which is free from technical
subtleties and permits an explicit representation
in terms of probability vectors as well as
illustration by means of concrete examples. We   
then change gears and explore the  Moran model with recombination. 
With the help
of standard techniques for Markov chains in continuous time,
we  develop the results stated above.

\section{The deterministic approach}
\label{sec:detmodel}

\subsection{The model and its recombinator formulation}

We start by extracting from  \cite{Baak05} and \cite{BaBa03}
the continuous-time dynamics of single 
crossovers in an infinite population. Let us consider genes or
chromosomes as 
linear arrangements of sites, indexed by the set $S:= \{ 0,1, \dots , n\}$;
sites may be interpreted as either nucleotide positions in a stretch of
DNA, or gene loci on a chromosome. For 
each site $i\in S$, there is a set $X_i$ of ``letters''  (to be interpreted
as nucleotides or alleles, respectively) that may possibly 
occur at that site. 
To allow for a convenient  notation, we restrict ourselves  to 
the simple but important case of \emph{finite} sets $X_i$; for the
full generality of arbitrary locally compact $X_i$, the reader is
referred to the original articles.

A \emph{type} is thus  defined 
as a sequence $x=(x^{}_0,x^{}_1, \dots , x^{}_n) \in X_0 \times X_1 \times \cdots 
\times X_n =:X$, where $X$ shall be called the \emph{type space}. By 
construction, $x^{}_i$ is the $i$-th coordinate of $x$, and we define 
$x^{}_I := (x^{}_i)_{i \in I}$ as the collection of coordinates 
with indices in $I$, where $I$ is a subset of $S$. Types 
may be understood as  alleles (if 
sites are nucleotide positions) or haplotypes (if sites are gene loci).
We shall think at the haploid level, speak of 
gametes as individuals, and describe a population as a distribution
of (absolute) frequencies
(or a non-negative measure) $\omega$ on $X$.
Namely, $\omega(\{x\})$ denotes the frequency of type $x \in X$, and
$\omega(A) := \sum_{x \in A} \omega(\{x\})$ for $A \subset X$;
we abbreviate $\omega(\{x\})$ as $\omega(x)$. The set of all
frequency distributions on $X$ is denoted by $\cM$.
If we define $\delta_x$ as the point measure on $x$ 
(i.e., $\delta_x(y)= \delta_{x,y}$ for $x,y \in X$), we can also write
$\omega = \sum_{x \in X} \omega(x) \delta_x$. We may,
alternatively, interpret
$\delta_x$ as the basis vector of $\RR_{\geq 0}^{\lvert X \rvert}$
that corresponds to $x$ (where a suitable ordering of types is implied,
and $\lvert X \rvert$ is the number of elements in $X$);
$\omega$ is thus
identified with a vector in $\RR_{\geq 0}^{\lvert X \rvert}$.


At this stage, frequencies need not be normalised;
$\omega(x)$ may simply be thought of as the size of the subpopulation
of type $x$, measured in units so large that it may be considered a
continuous quantity. 
The corresponding normalised version $p := \omega/\|\omega\|$ 
(where $\|\omega\|
:= \sum_{x \in X} \omega(x)=\omega(X)$ is the total population size)
is then a probability distribution on $X$, and may be identified
with a probability vector.

\begin{figure}[h]
  \psfrag{0}{$0$}
  \psfrag{1}{$1$}
  \psfrag{n}{$n$}
  \psfrag{iinS}{$i \in S$}
  \psfrag{12}{$\frac{1}{2}$}
  \psfrag{32}{$\frac{3}{2}$}
  \psfrag{ungerade}{$\frac{(2n-1)}{2}$}
  \psfrag{alpha}{$\alpha \in L$}
  \begin{center}
  \includegraphics{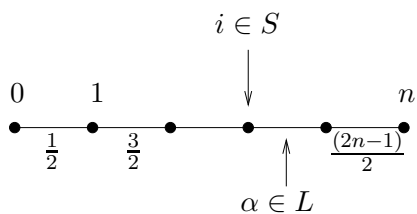}
  \end{center}
  \caption{\label{fig:sitesandlinks} {\small Sites and links.}}
\end{figure}

Recombination acts on the links 
between the sites; the links are collected
into the set $L := \left\{ \frac{1}{2} , 
\frac{3}{2}, \dots , \frac{2n-1}{2} \right\}$. We shall use Latin indices for 
the sites and Greek indices for the links, and the implicit rule will always 
be that $\alpha = \frac{2i+1}{2}$ is the link between sites $i$ and $i+1$;
see Figure~\ref{fig:sitesandlinks}.

\begin{figure}[h]
\psfrag{x}{$x^{}_0, \ldots , x^{}_n$}
\psfrag{y}{$y^{}_0, \ldots , y^{}_n$}
\psfrag{r}{$\varrho^{}_{\alpha}$}
\psfrag{xy}{$x^{}_0,\ldots,x^{}_{\lfloor \alpha \rfloor}, y^{}_{\lceil \alpha \rceil},\ldots, y^{}_n$}
\psfrag{yx}{$y^{}_0,\ldots,y^{}_{\lfloor \alpha \rfloor}, x^{}_{\lceil \alpha \rceil},\ldots, x^{}_n$}
\psfrag{*}{$* \;\,, \ldots , \; *$}
\psfrag{x*}{$x^{}_0,\ldots,x^{}_{\lfloor \alpha \rfloor}, 
            *\; , \; \ldots \; ,\; *$}
\psfrag{*x}{$* \;,\; \ldots \;, \; * \;, \; x^{}_{\lceil \alpha \rceil}, \ldots , x^{}_n$}
\hspace{-4cm}\includegraphics{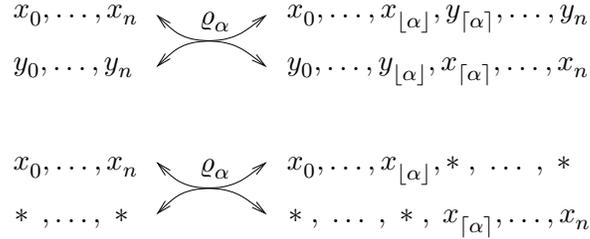}
\caption{\label{fig:reco} {\small 
Upper panel: Recombination between individuals of 
type $x$ and $y$. Lower panel:
The corresponding `marginalised' version that summarises all events
by which individuals of type $x$ are gained or lost.
Note that, in either case, the process can go both ways, as
indicated by the arrows.}}
\end{figure}

We shall make the simplifying assumption that only single crossovers occur at a
time (which is realistic if linkage is tight, i.e., if the sites belong 
to the DNA sequence of, say, less than a megabase, or represent a few 
adjacent loci). 
More precisely, for every $\alpha \in L$,
every individual exchanges, at rate $\varrho^{}_{\alpha}/2$, 
the sites after link $\alpha$ with those of a 
randomly chosen partner. Explicitly, if
the `active' and the partner individual are of type $x$ and $y$,
then the new pair has types $(x^{}_0,x^{}_1,\ldots,
x^{}_{\lfloor \alpha \rfloor}, y^{}_{\lceil \alpha \rceil}, \ldots, y^{}_n)$
and $(y^{}_0,y^{}_1,\ldots,
y^{}_{\lfloor \alpha \rfloor}, x^{}_{\lceil \alpha \rceil}, \ldots, x^{}_n)$,
where $\lfloor\alpha\rfloor 
(\lceil\alpha\rceil )$ is the largest integer below $\alpha$ (the 
smallest above $\alpha$);
see Fig.~\ref{fig:reco}. Since every individual
can occur as either the `active' individual or as its randomly
chosen partner, we have a total rate of  $\varrho^{}_\alpha$ for 
crossovers at link $\alpha$. 
For later use, we also define
$\varrho := \sum_{\alpha \in L} \varrho^{}_{\alpha}$.
The dynamics of the type frequencies is then given by the system of 
differential equations
\begin{equation} \label{eq:ode}
\dot{\omega}_t(x) =  \sum_{\alpha\in L} 
\varrho^{}_\alpha \big (\frac{1}{\|\omega_t\|} \omega_t (x^{}_0, \dots , 
x^{}_{\lfloor\alpha\rfloor}, \ast , \dots , \ast ) 
\omega_t (\ast , \dots , \ast ,
x^{}_{\lceil\alpha\rceil}, \dots ,x^{}_n) - \omega_t(x) \big )
\end{equation}
for all $x \in X$,
where  a ``$\ast$'' at site $i$ stands for
$X_i$ and
denotes marginalisation over the letters at site $i$
(i.e., $\omega_t (x^{}_0, \dots , x^{}_{\lfloor\alpha\rfloor}, \ast , \dots , \ast)$
is the frequency of individuals which have 
letters $x^{}_0, \dots , x^{}_{\lfloor\alpha\rfloor}$ at the sites before
$\alpha$, and arbitrary letters from $X_{\lceil \alpha \rceil}, \ldots,
X_n$ at the sites after $\alpha$).

Note that
the assumption of a reciprocal exchange (between two parents that combine
into {\em two} offspring under conservation of
the genetic material) is an `effective' description that results
from taking the symmetry of recombination into account. A more
detailed model starts out from two parents combining into {\em one}
offspring; but the symmetry of the process again leads to Eq.~\ref{eq:ode}
and is thus equivalent to our picture,  see
\cite[Ch.~II.2.1]{Buer00}. In contrast to the deterministic
situation, however, these details do play a role in finite populations;
this will be taken up in Sec.~\ref{sec:stoch}.

We have retained the non-normalised version for the frequencies here
for easier comparison with the stochastic model 
(the stochastic process will come naturally as a family 
of integer-valued random variables, which sum up to the population size
$N$).
 Of course,
the more familiar normalised version emerges if $\omega_t$
in \eqref{eq:ode} is replaced by $p_t := \omega_t / \| \omega_t \|$; 
the normalising factor on the right-hand side is then, of course, unity.

Note that 
the model \eqref{eq:ode} implies Hardy-Weinberg proportions
(i.e., frequencies of diplotypes are given by independent combination
of the corresponding haplotypes at all times). In 
continuous time, this only applies exactly if the duration
of the diplophase is negligible. However, it is a good approximation
if recombination is rare at the scale of an individual's lifetime.

An important ingredient to the solution of the
large, nonlinear  system of differential equations 
(\ref{eq:ode}) lies in its reformulation in terms of
recombination operators. Let us introduce the projection operators 
$\pi^{}_i$, $i \in S$,  via
\begin{equation}\label{eq:projector}
\begin{array}{rlcl}
\pi^{}_i : & X_0 \times X_1 \times \dots \times X_n & \to & X_i\\
& (x^{}_0, x^{}_1, \dots , x^{}_n) & \mapsto & x^{}_i,
\end{array}
\end{equation}
i.e., $\pi^{}_i$ is the canonical projection to the $i$-th coordinate.
Likewise, for any index set $I \subset S$, one defines a projector
$\pi^{}_I: X \to X_I := \btimes_{i \in I} X_i$  via 
$(x^{}_0,x^{}_1,\ldots, x^{}_n) \mapsto (x^{}_i)^{}_{i \in I} =: x^{}_I$. We shall frequently
use the abbreviations 
$\pi^{}_{<\alpha} := \pi^{}_{\{1,\ldots,\lfloor \alpha \rfloor\}}$ and
$\pi^{}_{>\alpha} := \pi^{}_{\{\lceil \alpha \rceil, \ldots, n\}}$,
as well as $x^{}_{<\alpha}:= \pi^{}_{<\alpha}(x)$, 
$x^{}_{>\alpha}:= \pi^{}_{>\alpha}(x)$. The projectors 
$\pi^{}_{<\alpha}$ and $\pi^{}_{>\alpha}$
may be thought of as \emph{cut and forget} operators 
because they take the leading or trailing segment of a sequence $x$, and 
forget about the rest. 

Whereas the $\pi^{}_I$  act on the types, 
we also need the induced mapping at the level of the population, namely,
\begin{equation} \label{eq:marg}
\begin{array}{rccl}
\pi^{}_I. : & \cM & 
           \longrightarrow & \cM \\
         &  \omega     & \mapsto & \pi^{}_I . \omega := \omega \circ \pi^{-1}_I,
\end{array}
\end{equation}
where $\pi^{-1}_I$ denotes the preimage under 
$\pi^{}_I$. 
The operation $.$ (where the dot is on the line and should not be
confused with a multiplication sign) is known as the ``pullback'' 
of $\pi^{}_I$ w.r.t. 
$\omega$; in terms of coordinates, the definition may be spelled out as
\[ 
  (\pi^{}_I . \omega)(x^{}_I) = \omega \circ \pi^{-1}_I (x^{}_I) 
  = \omega (\{x\in X \mid  \pi^{}_I(x) = x^{}_I \}), \quad 
  x^{}_I \in X_I.
\]  
So, $\pi^{}_I . \omega$ is
nothing but the marginal distribution of $\omega$ with respect
to the sites in $I$. In particular,  $(\pi^{}_{<\alpha} . \omega)(x^{}_{<\alpha})=
\omega (x^{}_0,x^{}_1, \ldots, x^{}_{\lfloor\alpha \rfloor}, \ast ,\dots ,\ast )$ is  
the marginal frequency  of  sequences prescribed at 
the sites before $\alpha$, and vice versa for the sites after $\alpha$. 
Note that, although we have
defined $\pi^{}_I .$ as acting on nonnegative measures (or vectors) only,
this linear operator  may be extended to the set of all measures.

Now, recombination (at the level of the population) means the relinking of 
a randomly chosen leading segment with a randomly chosen trailing segment.
We therefore introduce (elementary) recombination operators 
(or \emph{recombinators}, for short), $R_\alpha : \cM \to 
\cM$ for $\alpha \in L$, defined by 
\begin{equation}\label{eq:recombinator}
R_\alpha (\omega) := \frac{1}{\|\omega\|}
\big ( (\pi^{}_{<\alpha} . \omega) \otimes (\pi^{}_{>\alpha} . \omega) \big ).
\end{equation}
Here, the tensor product reflects the independent combination (i.e., the 
product measure) of the two marginals $\pi^{}_{<\alpha} . \omega$ and 
$\pi^{}_{>\alpha} . \omega$. $R_\alpha$ is therefore  a \emph{cut and relink}
operator. $R_\alpha (\omega)$ may be understood as the population 
that emerges if \emph{all} individuals of the population $\omega$ 
disintegrate into 
their leading and trailing segments, and these are relinked randomly;
the sites before $\alpha$ are then in linkage equilibrium with respect
to those after $\alpha$. Note that $\|R_\alpha (\omega)\| = \|\omega\|$.

In terms of coordinates, Eq.~\eqref{eq:recombinator} reads
\begin{equation}\label{eq:recombi_coord}
\big (R_\alpha (\omega) \big ) (x) = \frac{1}{\|\omega \|}
\omega (x^{}_0,x^{}_1, \dots , x^{}_{\lfloor\alpha\rfloor},
\ast ,\dots , \ast) \omega(\ast , \dots , \ast, x^{}_{\lceil\alpha\rceil},
\dots ,x^{}_n).
\end{equation}
The recombination dynamics (\ref{eq:ode}) may thus be compactly 
rewritten as
\begin{equation} \label{eq:compact}
\dot{\omega}_t = \sum_{\alpha\in L} 
\varrho^{}_\alpha \big ( R_\alpha (\omega_t) - \omega_t \big ) = 
\sum_{\alpha\in L} \varrho^{}_\alpha (R_\alpha -\one) (\omega_t)
=: \varPhi(\omega_t),
\end{equation}
where $\one$ is the identity operator.

\subsection{Solution via recombinators}

The solution of (\ref{eq:compact}) relies on some elementary properties of 
our recombinators. Most importantly, they are idempotents and commute
with each other, i.e.,
\begin{eqnarray} 
R^2_\alpha & = & R_\alpha , \qquad\  \alpha \in L, \label{eq:idpot}\\
R_\alpha R_\beta & = & R_\beta R_\alpha,  \quad \alpha , \beta \in L. 
\label{eq:commut}
\end{eqnarray}
These properties are intuitively very plausible: if linkage equilibrium has 
already been established at link $\alpha$, then further recombination 
at that link does not change the situation; and if a product measure is 
formed with respect to two links $\alpha$ and $\beta$, the result does not 
depend on the order in which the links are affected. For the proof, we 
refer  to \cite[Prop.\ 2]{BaBa03}; let us only 
mention here that it relies on the elementary fact that, for $\omega\in 
\cM$,
$$\begin{array}{ll}
\pi^{}_{<\alpha} . \big (R_\beta (\omega) \big ) = \pi^{}_{<\alpha} . \omega \quad & 
\mbox{ for } \beta  \geq \alpha , \mbox{ and} \\
\pi^{}_{>\alpha} . \big (R_\beta (\omega) \big ) = \pi^{}_{>\alpha} . \omega & \mbox{ for } 
\beta \leq \alpha ;
\end{array}
$$
that is, recombination at or after $\alpha$ does not affect 
the marginal frequencies
at sites before $\alpha$, and vice versa.

We now define \emph{composite} recombinators as
$$
R_G := \prod_{\alpha\in G} R_\alpha \qquad \mbox{ for } G\subset L.
$$
Here, the product is to be read as composition; it is, indeed, a
matrix product if the recombinators are written in their matrix
representation, which is available in the case of finite types
considered here (see \cite{Baak01}).
By property (\ref{eq:commut}), the order in the composition plays no role.
Furthermore, (\ref{eq:idpot}) and (\ref{eq:commut}) obviously entail 
$R_G R_H = R_{G\cup H}$ for $G,H \subset L$.

The solution of the single-crossover dynamics 
(\ref{eq:compact}) may now be given in closed form as
\begin{equation}\label{eq:recosol}
\omega_t = \sum_{G\subset L} a_G (t) R_G (\omega_0) =: \varphi_t(\omega_0)
\end{equation}
with coefficients 
\[
   a_G (t) = \prod_{\alpha\in L\backslash G} 
e^{-\varrho^{}_\alpha t} \prod_{\beta\in G} (1-e^{-\varrho^{}_\beta t}),
\] 
and 
initial value $\omega_0$; i.e., $\varphi_t$ is the {\em semigroup}
belonging to the recombination equation \eqref{eq:compact}.
For the proof,  the reader is referred to
\cite[Thm.\ 2]{BaBa03}, or \cite[Thm.\ 3]{Baak05} (the former article
contains the original, the latter 
a shorter and more elegant version of the proof).  
Let us only note here that the coefficients 
$a_G(t)$ have the following intuitive 
explanation. In a given individual, consider the set of links that has, 
until $t$, been hit by at least one crossover event. The probability that 
this set is $G$ equals the  probability that all links in $G$ 
have already been hit, while those in the complement of $G$ have not; by independence across links, this probability is just 
$a_G(t)$. 

We would like to note, however, that this plausible argument should
not be taken too far: For example, an analogous solution, although
suggestive, does {\em not} apply to the corresponding single-crossover
model in {\em discrete} time (its solution is, in fact, much more
difficult, see the discussion in \cite{BaBa03}). Indeed, 
explicit solutions to  large, nonlinear  systems are  rare
 -- explicit semigroups are usually available for linear systems
at best. For this reason, 
the recombination equation
and its solution have already been taken up in the framework of
functional analysis, where they have led to an extension of potential
theory \cite{Popa07}. In the next paragraph, we will meet some further
clues to an underlying linear structure that lurks behind the solution.

\subsection{Linkage disequilibria}

The approach described in the previous Section  is somewhat unconventional
from a population genetics perspective,
in that it solves the recombination dynamics at the level of the type
frequencies. As mentioned in the Introduction, however, 
one usually works with transformed 
quantities, like moments, cumulants, or linkage disequilibria (LDEs). 
Indeed, certain linkage 
disequilibria are intimately connected with the solution just presented, and 
provide the key to an underlying linearity, and hence simplicity, as we shall 
now see. Following \cite{BaBa03}, let us define what we shall
call {\em LDE operators} via
\begin{equation}\label{eq:T_G}
T_G := \sum_{{\text{\d{$H$}}} \supset G} (-1)^{|H\setminus G|} R_H, \qquad G\subset L, 
\end{equation}
where the underdot indicates the summation variable. 
Eq.~(\ref{eq:T_G}) leads to the inverse $R_G = \sum_{{\text{\d{$H$}}}
\supset G} T_H$ by M\"obius inversion,  a versatile tool from 
combinatorial theory, see \cite[Thm.~4.18]{Aign79}.
It was shown 
in \cite{BaBa03} that, if $\omega_t$ is the solution \eqref{eq:recosol}, 
the transformed quantities $T_G(\omega_t)$ satisfy
\begin{equation}\label{eq:TG_decay}
\frac{d}{dt} T_G (\omega_t) = - \left( \sum_{\alpha \in L \setminus G} 
\varrho^{}_\alpha \right) 
T_G (\omega_t), \qquad G\subset L, 
\end{equation}
which is a system of \emph{decoupled, linear, homogeneous} differential equations, with solution 
$T_G (\omega_t) = \exp(-\sum_{\alpha \in L \setminus G} \varrho_{\alpha})
T_G (\omega_0)$.
Note that this simple form emerged through the \emph{nonlinear} transform~(\ref{eq:T_G}) as applied to the solution of the \emph{coupled, nonlinear} 
differential equation \eqref{eq:compact}.

Obviously, the $T_G(\omega_t)$ provide candidates for the definition of linkage 
disequilibria which decouple, and decay exponentially
(unless, of course, $G=L$, which corresponds to the stationary state
\begin{equation}\label{eq:stat_state}
  \omega_{\infty}=T_L(\omega_0) = \frac{1}{\|\omega\|^{n-1}}
  \bigotimes_{i =1}^{n} (\pi^{}_{i} . \omega_0),
\end{equation}
the population in linkage equilibrium with respect to
all links). 
All that remains to be done is to identify a set of suitable components to work 
with (using all components of $T_G(\omega_t)$, for all $G\subset L$, would imply
a lot of redundancy).

To this end, let us introduce the following shorthand notation. Let $\langle j_1,
\ldots , j_k\rangle$, with $j_1< \dots <j_k$, symbolically denote a 
so-called {\em cylinder set} in 
$X$ that is specified at sites $j_i \in S$, for $1 \leq i \leq k$. More precisely, 
these are sets of the form
\begin{equation}
\langle j_1, \dots ,j_k\rangle = X_0 \times \dots \times X_{j_1-1} \times 
\{ x^{}_{j_1} \} \times [ \dots ] \times \{ x^{}_{j_k} \} 
\times X_{j_{k+1}} \times \dots 
\times X_n, \label{eq:prec}
\end{equation}
where $[\dots ]$ contains factors $\{ x^{}_i \}$ or $X_i$ depending on whether $i$ 
appears in the set $\{ j_1, \dots ,j_k \}$ or not. That is, with the cylinder 
set $\langle j_1, \dots ,j_k\rangle$, we mean the set of types that have
prescribed 
letters $x^{}_{j_1},\dots ,x^{}_{j_k}$ at sites $j_1,\dots ,j_k$, and arbitrary letters 
at all other sites.  
Note that the shorthand $\langle j_1,\dots ,j_k\rangle$ is 
symbolic in that it does not specify the letters explicitly (the $x^{}_{j_i}$ appear 
on the right-hand side of (\ref{eq:prec}), but not on the left). Note also
that the cylinder sets formalise our previous `$*$' notation:
$\omega_t(\langle j_1, \dots ,j_k\rangle)$ is the marginal frequency
of the types prescribed at sites $j_1, \ldots, j_k$, and \eqref{eq:ode} may be
rewritten as
\[
\dot \omega_t(\langle S \rangle) = \sum_{\alpha \in L} \varrho^{}_{\alpha} 
 \big ( \frac{1}{\| \omega_t \|}
 \omega_t(\langle 0,\ldots,\lfloor \alpha \rfloor \rangle) \,
\omega_t(\langle \lceil \alpha \rceil, \ldots, n \rangle) 
- \omega_t(\langle S \rangle ) \big ).
\]

Let us now define \emph{linkage disequilibria} (or {\em correlation
functions})  of order $k$ via 
\begin{equation}
d_t \left( \langle j_1 ,\dots ,j_k \rangle\right) 
:= \left( T_{\{ \alpha <j_1 \} 
\cup \{ \alpha >j_k\}} (p_t) \right) 
\left( \langle j_1,\dots ,j_k\rangle\right). 
\label{eq:d_t}
\end{equation}
That is, for a given cylinder set, LDEs emerge by applying, to 
$p_t = \omega_t/\|\omega_t\|$, the LDE operator
defined by the links before the first and after the last element
of that cylinder set, and evaluating the resulting quantity
at the cylinder set.  Clearly, $d_t \left( \langle j_1 ,\dots ,j_k \rangle\right)$
contains products of at most $k$ marginal frequencies. Note that we have
defined linkage disequilibria on the basis of the {\em normalised} quantities
$p_t$ for the sake of compatibility, and comparability,
with related quantities in population genetics.

By \eqref{eq:TG_decay} we know that, for every cylinder set
$\langle j_1, \dots ,j_k\rangle$,
\begin{equation}\label{eq:decay}
\frac{d}{dt} d_t  ( \langle j_1, \dots ,j_k\rangle  ) = -\Big( \sum_{j_1<
 \alpha <j_k}
\varrho^{}_\alpha \Big ) d_t  ( \langle j_1,\dots ,j_k\rangle ) ;
\end{equation}
we thus have linkage disequilibria \emph{of all orders} that decouple and decay 
exponentially. It was shown in \cite[p.~25]{BaBa03} that the collection 
of $d_t (\langle 
j_1,\dots ,j_k\rangle)$, for all index sets $\{ j_1, \dots , j_k \} \subset S$ and all 
choices of letters $x^{}_{j_i} \in X_{j_i}, 1 \leq i \leq k$, is complete in that it 
uniquely determines $p_t$. 

{\em An example.} Consider $S = \{ 0,1,2,3\}$. The highest (fourth-order) 
LDE is, on the basis of (\ref{eq:T_G}) 
and (\ref{eq:d_t}), given by
\[
\begin{split}
d_t ( \langle 0,1,2,3\rangle ) 
= &\big( T_\varnothing (p_t) \big) (\langle 0,1,2,3\rangle)
= \sum_{{\text{\d{$H$}}}\subset L} (-1)^{|H|} \big (R_H (p_t) \big ) 
(\langle 0,1,2,3\rangle) \\
 =  & [ 0,1,2,3 ] - [ 0 ] [ 1,2,3 ] -  [ 0,1 ] [ 2,3 ] -  [0,1,2][ 3 ] \\
 +  & [ 0 ] [ 1 ] [ 2,3 ] + [ 0 ][ 1,2 ] [ 3 ]   + [ 0,1] [ 2][ 3 ] \\
 -  & [ 0][1][2][3],
\end{split}
\]
where we have now used the shorthand $[j_1, \ldots, j_k]:= 
p_t(\langle j_1, \ldots, j_k \rangle )$. By \eqref{eq:decay},
we have $\dot d_t( \langle 0,1,2,3\rangle) = 
-(\varrho^{}_{1/2}+\varrho^{}_{3/2}+\varrho^{}_{5/2}) d_t( \langle 0,1,2,3\rangle)$.

The third-, second-, and first-order LDEs are given by 
\begin{eqnarray*}
d_t(\langle j_1,j_2,j_3 \rangle) &=&[j_1,j_2,j_3] 
- [ j_1 ] [ j_2,j_3 ]
- [ j_1,j_2 ] [  j_3 ]
+ [ j_1 ] [ j_2 ] [ j_3 ], \\
d_t(\langle j_1,j_2 \rangle) &= & [ j_1,j_2 ] - [ j_1 ][ j_2 ], \\
d_t(\langle j_1 \rangle ) &= & [ j_1 ],
\end{eqnarray*}
where the latter correspond to the letter frequencies, as usual.

Obviously, the $d_t$ are correlation functions that measure
the dependence between sites, at the various orders.
It is important to note that 
they agree with other measures of LDE 
(see \cite[Ch.~V.4.2]{Buer00} for an overview) only up to order two.
From order three onwards, they are special in that 
terms like $[0,2][1]$ or $[0,3][1,2]$ are absent. This is due 
to the fact that our $d_t$
are based on \emph{ordered} partitions of $S$, and thus only contain
terms `produced' by composite recombinators; they are
therefore adapted to single-crossover dynamics.
In contrast,
conventional LDEs or central moments are based on
{\em unordered} partitions, which cannot be produced by composite 
recombinators.

%
%
\section{The stochastic model and its simulation}
\label{sec:stoch}

The finite population  counterpart of our deterministic model 
is the Moran model with single-crossover
recombination. To simplify matters (and in order to clearly dissect the 
individual effects of recombination and resampling),
we shall assume that resampling (traditionally
referred to as genetic drift) and recombination occur independently of
each other. More precisely, we assume
a finite population of fixed size $N$, in which every individual 
experiences, independently of the others, 
\begin{itemize}\itemsep-1ex
\item resampling  at rate $b/2$. The individual reproduces, 
the offspring inherits the parent's type and replaces a
randomly chosen individual (possibly its own parent).\\
\item recombination  at (overall) rate $\varrho^{}_{\alpha}$ at link $\alpha \in L$. 
Every individual picks  a random partner (maybe itself) at rate 
$\varrho^{}_{\alpha}/2$, and the pair exchanges
the sites after link $\alpha$. That is,  if the recombining individuals
have types $x$ and $y$, they are replaced by the two offspring
individuals
$(x^{}_{<\alpha},y^{}_{>\alpha})$ and $(y^{}_{<\alpha},x^{}_{>\alpha})$, as in 
the deterministic case, and Fig.~\ref{fig:reco}. That is, the genetic
material is conserved, and any resampling effect is excluded.
As before,
the per-capita rate of
recombination at link $\alpha$ is then $\varrho^{}_{\alpha}$, because
both orderings of the individuals lead to the same event.
To avoid degeneracies, we shall assume throughout that 
$\varrho^{}_{\alpha}>0$ for all $\alpha \in L$.
\end{itemize}

As in the deterministic model, the factor $1/2$ arises
every time {\em ordered} pairs (of an active individual and a randomly chosen 
partner) are considered (it is the same factor that also occurs when the Moran
and Wright-Fisher models are compared \cite[p.~23]{Durr02}).
Note that the randomly chosen second individual
(for resampling or recombination)
may be the active individual itself; then, effectively, nothing happens.
One might, for biological reasons, prefer to exclude these events by
sampling from the remaining population only; but this means nothing but
a change in time scale of order $1/N$.

To formalise this verbal description of the process,
let the state of the population at time $t$ be given by the collection
$Z_t = \big(Z_t(x)\big)_{x\in X} \in 
E:=\{z \in \{0,1,...,N\}^{|X|}_{} \mid \sum_{x} z(x) = N\}$, where
$Z_t(x)$ is the number of individuals of type $x$ at time $t$; clearly, 
$\sum_{x\in X}Z_t(x)=N$. We  also use $Z_t$ in the sense of a
(random counting) measure, in analogy with $\omega_t$ (but keep
in mind that $Z_t$ is integer-valued and counts single individuals,
whereas $\omega_t$
denotes continuous frequencies in an infinite population).
The letter $z$ will be used to denote realisations of $Z_t$ --- but 
note that the symbols $x,y$, and $z$ are not on equal
footing ($x$ and $y$ will continue to be types).
The stochastic process $\{Z_t\}_{t \geq 0}$ is the continuous-time Markov
chain on $E$ defined as follows.
If the current state is $Z_t=z$, two types of  transitions may occur:
\begin{eqnarray} 
&\text{resampling:} & z \rightarrow z + s(x,y), \quad 
                      s(x,y):= \delta_x - \delta_y  \nonumber, \\
 && \text{at rate } \frac{1}{2N} b z(x) z(y) \; \text{ for }  
    (x,y) \in X \times X \label{eq:rate_resampling}\\
&\text{recombination:} & z \rightarrow z +r(x,y,\alpha), \nonumber \\
&                      & r(x,y,\alpha) :=  
\delta_{(x^{}_{<\alpha},y^{}_{>\alpha})} +\delta_{(y^{}_{<\alpha},x^{}_{>\alpha})} - \delta_x - \delta_y, \nonumber \\
 &&\text{at rate } \frac{1}{2N}
    \varrho^{}_{\alpha}z(x) z(y) \; \text{ for } 
  (x,y) \in X \times X, \alpha\in L \label{eq:rate_reco}
\end{eqnarray}
(where $\delta_x$ is the point measure on $x$, as before).
Note that, in \eqref{eq:rate_resampling}
and \eqref{eq:rate_reco}, transitions that leave $E$ are 
automatically excluded by the fact that  the corresponding rates vanish.

Note that `empty transitions' ($s(x,y)=0$ or $r(x,y,\alpha)=0$) are 
explicitly included 
(they occur  if $x=y$ in  resampling
or recombination, and if $x^{}_{<\alpha}=y^{}_{<\alpha}$ or
$x^{}_{>\alpha}=y^{}_{>\alpha}$ in recombination). This is convenient since
the total reproduction and 
recombination rates in the population (as based on active individuals)
are then given by $bN/2$ and $\varrho N/2$, 
respectively, independently of $z$, a fact that comes in handy in the 
simulations. Since we only require $Z_t$ at (equidistant) epochs 
$t_i=i\Delta t$, $i =0,1,2,\ldots,$ no waiting times for individual events 
need be generated;
rather, we may draw the total number of events in a time interval
$\Delta t$ from the Poisson distribution with parameter 
$N(b+\varrho)\Delta t/2$. The nature of each
event is then determined in the obvious way (the active individual is of 
type $x$ with probability $z(x)/N$; the second individual (chosen
 randomly for resampling or recombination)
is of type $y$ with probability $z(y)/N$; the pair performs resampling
with probability $b/(b+\varrho)$, or recombination with probability 
$\varrho/(b+\varrho)$; if recombination occurs, then at link $\alpha$ with 
probability $\varrho^{}_{\alpha}/\varrho$). 

For simulations, the description of the process through 
\eqref{eq:rate_resampling} and \eqref{eq:rate_reco} is
perfectly adequate. For the theoretical analysis, however,
one also needs the rates $Q(z,z+v)$ for the transitions $z \to z+v$
for {\em given} $z \in E$, $v \in E-z$ 
(where $E-z := \{v \mid z+v \in E\}$). 
Here, one must take
care of the fact that distinct combinations of $x,y$, and $\alpha$
may lead to the same `net' transition. 
Clearly, $Q(z,z+v)=Q^{(s)}(z,z+v)+Q^{(r)}(z,z+v)$,
where
\begin{eqnarray}\label{eq:Q_s}
 Q^{(s)}(z,z+v)& := &  \frac{b}{2N} \sum_{\substack{(x,y) \in X \times X:\\
                  s(x,y)=v}}  z(x) z(y), \\
 \label{eq:Q_r}
 Q^{(r)}(z,z+v)& := &  \frac{1}{2N} \sum_{\substack{(x,y) \in X \times X, \, \alpha \in L:\\
                  r(x,y,\alpha)=v}}  \varrho^{}_{\alpha} z(x) z(y). 
\end{eqnarray}
Of course, the empty sum 
corresponds to impossible transitions and is understood as $0$. Note that the sum in \eqref{eq:Q_s}
contains more than one term only in the case $v=0$; however, 
\eqref{eq:Q_r} is a `true' sum (of at least two terms) whenever 
$v$ is of the form $r(x,y,\alpha)$ for some $x,y,\alpha$.

Note that, with the  rates for `empty transitions'
as ascribed to the diagonal elements $Q(z,z)$,
the collection $(Q(z,z'))_{z,z' \in E}$ does {\em not} form a
Markov generator; but the corresponding generator 
$\widetilde Q=(\widetilde Q(z,z'))_{z,z' \in E}$ is obtained by defining
$\widetilde Q(z,z') = Q(z,z')$, for $z'\neq z$, together with
$\widetilde Q(z,z) = - \sum_{z': z' \neq z} Q(z,z')$.

This rather informal definition of the Markov chain in terms of its transitions
and
their rates is all that is required for  this article -- after all,
we are in the well-behaved situation of a finite state space.  
Excellent overviews of many aspects of Markov chains
in  continuous time (including the formal details)
can  be found in \cite[Ch.~2]{Norr97} or \cite[Ch.~I.8]{Assm03}.

In contrast to the situation in infinite populations, it now matters
whether one or two offspring are created at a time. For comparison, we
shall, therefore, also consider the 
alternate (biologically more realistic) recombination scheme in which
 an individual recombines with a random partner,
but only {\em one} of the offspring
is kept (chosen randomly), and replaces {\em one} of its  parents 
(again chosen randomly).
That is, instead of \eqref{eq:rate_reco}, we now have four recombination
transitions
\begin{eqnarray} 
\label{eq:trans1_oneoff}
& z \rightarrow z - \delta_x  +\delta_{(x^{}_{<\alpha}, y^{}_{>\alpha})}, \quad 
& z \rightarrow z - \delta_y  +\delta_{(y^{}_{<\alpha},x^{}_{>\alpha})}, 
\\ \label{eq:trans2_oneoff}
& z \rightarrow z - \delta_x  +\delta_{(y^{}_{<\alpha},x^{}_{>\alpha})}, \quad  & z \rightarrow z - \delta_y  +\delta_{(x^{}_{<\alpha},y^{}_{>\alpha})}, 
\end{eqnarray}
each at rate
\begin{equation}\label{eq:rate_oneoff}
 \frac{1}{8N} \varrho^{}_{\alpha}z(x) z(y),  
\text{ for } (x,y) \in X\times X, \alpha\in L.
\end{equation}
Note that, in contrast to \eqref{eq:rate_reco}, the symmetry between $x$
and $y$ is broken in single transitions. Furthermore, only half as many
replacements happen here;
the corresponding deterministic dynamics is,
therefore, given by \eqref{eq:compact} with $\varrho^{}_{\alpha}$ replaced
by $\varrho^{}_{\alpha}/2$ for all $\alpha \in L$.
More importantly, the recombination scheme \eqref{eq:trans1_oneoff},
\eqref{eq:trans2_oneoff}
no longer conserves the genetic material 
(since net replacements (e.g.\ of $x^{}_{>\alpha}$
 by $y^{}_{>\alpha}$) take place), which introduces
a (minor) resampling effect not present in 
\eqref{eq:rate_reco}. If not stated otherwise, however,
we will stick to \eqref{eq:rate_reco} for our recombination transitions.

\section{Connections between stochastic and deterministic models}

Let us now explore the connection between the stochastic process 
$\{Z_t\}_{t\geq 0}$ on $E$, its normalised version 
$\{\widehat Z_t\}_{t\geq 0}=\{Z_t\}_{t\geq 0}/N$ on $E/N$,
and the solution $\omega_t=\varphi_t(\omega_0)$ of
the differential equation \eqref{eq:recosol}. 
We shall first consider the case without resampling
(i.e., $b=0$) and show that $\EE(Z_t)=\varphi_t(Z_0)$ (and, consequently,
$\EE(\hZ_t)=\varphi_t(\hZ_0)$), for {\em arbitrary} $N$.
For $b>0$, still with finite $N$, this is no longer true; 
however, $\hZ_t$ converges to $\varphi_t(p_0)$ for $N \to \infty$,
provided $\hZ_0$ converges to $p_0$. We 
shall make this convergence explicit, and 
investigate it by means of simulations.

\subsection{General properties of the Moran model with recombination}
For lack of reference and the sake of completeness, 
we start by making explicit a plausible -- and often implicitly-used -- 
fact concerning the evolution of the expectation  
in finite-state continuous-time Markov chains.

\begin{fact}
\label{fact:exp_change}
Let $\{Z_t\}_{t\geq 0}$ be a continuous-time Markov chain on 
a finite subset $E$ of $\ZZ^d$, as defined by its transitions
from state $z$ to state $z+v$ ($z, z+v \in E$) taking
place at rates $Q(z,z+v)$. For all $t \geq 0$,
the expectation of $Z_t$ then satisfies the equation
\begin{equation}\label{eq:ivp}
\frac{d}{dt}\mathbb{E}(Z_t) = \EE \big(F(Z_t)\big), 
\end{equation}
where, for $z \in E$, 
\begin{equation} \label{eq:F_def}
F(z) := \sum_{v \in E-z}\, v Q(z,z+v).
\end{equation}
\end{fact}
  
Before turning to the proof, let us note that $F(z)$ may be interpreted
as the `mean rate-of-change vector' of the chain in state $z$.
A little later, we shall turn \eqref{eq:ivp} into a
differential equation for $\EE(Z_t)$.

\begin{proof}
Let us first recall  the elementary fact that, for the Markov chain
starting in $Z_0=s$, we have $\PP(Z_t=z) = P_t(s,z)$,
where we have suppressed dependence on $s$ on the left-hand side,
and $P_t = (P_t(s,z))_{s,z \in E}= e^{t \widetilde Q}$ 
is the Markov semigroup corresponding to $\widetilde Q$,
the  generator of the chain (obtained from the collection of
transition rates $(Q(z,z+v))_{z,z+v \in E}$, see above). 
Therefore, $\EE(Z_t) = \sum_{z \in E} z P_t(s,z)$, and,
since $\frac{d}{dt} P(t) = P(t) \widetilde Q$, 
\begin{eqnarray*}
\frac{d}{dt} \EE (Z_t) & = & \sum_{z' \in E} z' \frac{d}{dt} P_t(s,z') 
 =  \sum_{z,z' \in E}  z' P_t(s,z) 
      \widetilde Q(z,z') \\
& = & \sum_{z,z' \in E}  (z'-z) P_t(s,z)
      \widetilde Q(z,z') 
 =  \sum_{z \in E} \; \sum_{v \in E-z} v P_t(s,z) \widetilde Q(z,z+v) \\
& = & \EE \big ( \sum_{v \in E-Z_t} v \widetilde Q(Z_t, Z_t + v) \big ) 
 =  \EE \big ( F(Z_t) \big ),
\end{eqnarray*}
where the third equality sign is due to the fact that $\widetilde Q$
is a Markov generator.
\end{proof}

After this digression, let us return
to the Moran model (with recombination {\em and} resampling), and
show that  $F=\varPhi$ on $E$, where $\varPhi$ continues to be the right-hand
side of \eqref{eq:compact}.

To this end, recall that $Z_t$ is a (counting) measure,
to which our projection operators may be applied in the usual way. 
In agreement with \eqref{eq:marg}, we write
$
\pi^{}_I . Z_t = Z_t \circ \pi_I^{-1} ,
$
i.e., $\pi^{}_I . Z_t$ is the marginal of $Z_t$ w.r.t.\ 
the sites in $I$; and, likewise, for a realisation $z$ of $Z_t$. 
Again, we also use shorthands such as $Z_t(x^{}_{<\alpha},*)$ to denote
marginal frequencies (in this case, $(\pi^{}_{<\alpha}.Z_t)(x^{}_{<\alpha})$). 
Furthermore, we set $E_I :=  \pi^{}_I . E$ 
for $I \subset S$.

To calculate $F$, let us write
$F=F^{(s)} + F^{(r)}$, where $F^{(s)}$
and $F^{(r)}$ 
take care of transitions due to resampling and recombination,
respectively. By symmetry of the transition rates for resampling,
$F^{(s)}=0$ (type frequencies are 
martingales under resampling alone); we can therefore restrict
ourselves to the recombination transitions defined by $Q^{(r)}(z,z+v)$
of \eqref{eq:Q_r}.
Summing, as in \eqref{eq:Q_r}, over all possibilities for the gain or 
loss of a single
$x$-individual (cf.~Fig.~\ref{fig:reco}, lower panel),
one obtains for the $x$-component of $F$ 
\begin{equation}
\begin{split}
\label{eq:F=Phi}
F_x(z) & = F^{(r)}_x(z)  = 
\sum_{\substack{v \in E-z:\\ v(x)=\pm 1}} v(x) Q^{(r)}(z,z+v)\\
& =  \sum_{\alpha \in L} \frac{1}{N} \varrho^{}_{\alpha} 
\big ( z(x^{}_{<\alpha},*) z(*,x^{}_{>\alpha}) 
- z(x^{}_{<\alpha},x^{}_{>\alpha})  z(*,*) \big ),
\end{split}
\end{equation}
where, of course, $z(x^{}_{<\alpha},x^{}_{>\alpha})=z(x)$, and $z(*,*)=N$.
One immediately recognises the familiar structure of \eqref{eq:ode}. We
may therefore conclude:
\begin{fact}
\label{fact:F=Phi}
In the Moran model with resampling and recombination 
transitions according to \eqref{eq:rate_resampling} and \eqref{eq:rate_reco},
we have $F=\varPhi$ on $E$,
with $\varPhi$ of \eqref{eq:compact}.
\end{fact}

\subsection{Recombination without resampling: expectations for finite $N$}

Let us now return to Eq.~\eqref{eq:ivp}. 
Note that, in general, 
it does not lead to a ``closed'' 
differential equation for $\mathbb{E} (Z_t)$,  because
it is not clear whether $\EE(F(Z_t))$ can be written as a function
of $\EE(Z_t)$ alone. 
Clearly, $\EE(F(Z_t))=F(\EE(Z_t))$ if $F$ is linear (or affine), as, 
for example, in 
Markov branching processes, or in the Moran model with mutation, but without
recombination. 
But for nonlinear $F$, this tends to be violated, as  nicely
illustrated in \cite[Ch.~1.4]{HJV05} for the case of the Ricker model
in ecology. If $F$ 
is nonlinear but polynomial (the usual case in population genetics
or chemical reaction systems, for example),
(\ref{eq:ivp}) can still 
serve as a starting point for an expansion involving a hierarchy of 
moments, which can  be closed by suitable approximation methods
(like, for example, the quasi-linkage equilibrium approach in
\cite{BaTu91}).

Our aim in this paragraph is
to show that recombination {\em without} resampling  
is another (and apparently new) exception: despite its nonlinearity, 
$\varPhi$ satisfies 
$\mathbb{E} (\varPhi(Z_t)) = \varPhi(\mathbb{E} (Z_t))$, 
provided the marginals of $Z_0$ are independent,
which always applies if $Z_0$ is fixed. This will require
a lemma concerning the independence of marginal processes.
For $I \subset S$, we introduce the `stretch' 
\[
  J(I) := \{i \in S \mid \min(I) \leq i \leq \max(I)\},
\]
and look at the projection of the recombination process on non-overlapping
stretches. This is the content of the following lemma.

\begin{lemma}\label{lem:indep_marginals} 
Let $\{Z_t\}_{t\geq 0}$ be the recombination process without resampling 
as defined by the transition rates \eqref{eq:rate_reco}. Let $A,B \subset S$ with
$J(A) \cap J(B)=\varnothing$.
Then, $\{\pi^{}_A . Z_t\}_{t \geq 0}$ and $\{\pi^{}_B . Z_t\}_{t \geq 0}$
are conditionally (on $Z_0$) independent 
Markov chains on $E_A$ and $E_B$.
\end{lemma}

\begin{proof}
Clearly, for any given $I \subset S$, $\{\pi^{}_I . Z_t\}_{t\geq 0}$ is a 
stochastic 
process on $E_I$. Let us first show that $\pi^{}_A . Z_t$ and
$\pi^{}_B . Z_t$ are individually Markov chains, and then
establish that they are (conditionally) independent.
For the first step, observe that, when $Z_t$ performs the transition
$z \to z+v$ (on $E$), with
$v= \delta_{(x_{<\alpha}, y^{}_{>\alpha})} 
+  \delta_{(y^{}_{<\alpha}, x^{}_{>\alpha})}- \delta_x - \delta_y  $
for some
$\alpha \in L$ and  $x,y \in X$, 
then $\pi^{}_I . Z_t$ 
goes from $\pi^{}_I .z \to \pi^{}_I . (z +v)$ on $E_I$, where 
$\pi^{}_I . v = 
\delta_{\pi^{}_I(x^{}_{<\alpha}, y^{}_{>\alpha})} 
+ \delta_{\pi^{}_I(y^{}_{<\alpha}, x^{}_{>\alpha})}
 - \delta_{\pi^{}_I(x)} - \delta_{\pi^{}_I(y)}$.
The rate for a corresponding `projected transition' $v^{}_I \in E_I-z_I$ 
($z_I \in E_I$) is then given by summing all rates
of the original process that lead to the transition $v^{}_I$ under the
projection. That is, with $r^{}_I(x^{}_I,y^{}_I,\alpha) :=
\delta_{(x^{}_{I_{<\alpha}},y^{}_{I_{>\alpha}})} 
+ \delta_{(y^{}_{I_{<\alpha}},x^{}_{I_{>\alpha}})} 
-\delta_{x^{}_I} - \delta_{y^{}_I}$ and the shorthand
$I_{<\alpha} := I \cap \{0,1,\ldots, \lfloor \alpha \rfloor \}$
(and likewise for $I_{>\alpha}$),
one gets
\[
\sum_{v: \pi^{}_I . v = v^{}_I} Q^{(r)}(z,z+v) =  
   \frac{1}{2N} \sum_{\substack{x^{}_I,y^{}_I\in X_I, \,  \alpha \in L:\\
    r^{}_I(x^{}_I,y^{}_I,\alpha)  = v^{}_I}}  \varrho_{\alpha}
   \big ( (\pi^{}_I . z)(x^{}_I) \big )  \big ( (\pi^{}_I . z)(y^{}_I) \big ), 
\]
for $z \in E$.
Since these rates only depend on $\pi^{}_I . z$ (rather than $z$ itself), 
they define a collection of rates 
$Q_I^{(r)}(z^{}_I,z^{}_I+v^{}_I)$ ($z^{}_I \in E_I, v^{}_I \in E_I-z_I$) so that
$\sum_{v: \pi^{}_I . v = v^{}_I} Q^{(r)}(z,z+v) 
= Q_I^{(r)}(\pi^{}_I.z, \pi^{}_I.(z+v))$ for all $z \in E$.
Therefore, the marginalised process  $\{\pi^{}_I.Z_t\}_{t\geq 0}$ is a 
Markov chain on $E_I$ with transitions
$z^{}_I \to z^{}_I + v^{}_I$ at rates $Q_I^{(r)}(z^{}_I,z^{}_I+v^{}_I)$.
(This is an example of the so-called {\em lumping procedure} for
Markov chains, compare \cite{BuRo58} and \cite{KeSn81} for the
general case, or \cite{BBBK05} for the sequence context considered
here).
Our processes $\{\pi^{}_A . Z_t\}_{t \geq 0}$ and $\{\pi^{}_B . Z_t\}_{t \geq 0}$ are therefore 
Markov on $E_A$ and $E_B$, respectively.

For the second step, we note that,  
for  given $\alpha$ and $I$, a net transition in $\pi^{}_I.Z_t $ (i.e., $\pi^{}_I.v \neq 0$)
requires $\lfloor \alpha \rfloor, \lceil \alpha \rceil \in J(I)$.
Since $J(A) \cap J(B)=\varnothing$ by assumption,
$\pi^{}_A. v \neq 0$ implies $\pi^{}_B.v = 0$ and vice versa. 
A transition in 
$\{\pi^{}_A . Z_t\}_{t \geq 0}$ therefore leaves $\{\pi^{}_B . Z_t\}_{t \geq 0}$ unchanged, and
vice versa. The joint process 
$W_t := \{\pi^{}_A . Z_t,\pi^{}_B . Z_t\}_{t \geq 0}$ therefore has 
generator 
$\tilde Q_{A,B} = \tilde Q_A \otimes \one_B + \one_A \otimes \tilde Q_B$,
where $\tilde Q_A$ is the generator of 
$\{\pi^{}_A . Z_t\}_{t \geq 0}$, $\tilde Q_B$  the generator of
$\{\pi^{}_B . Z_t\}_{t \geq 0}$, and $\one_A$ and $\one_B$ denote
the identity on $E_A$ and $E_B$, respectively.
Hence, the marginal processes are 
conditionally independent, and the claim
follows. 
\end{proof}

\begin{remark}\label{rem:mutual}
Although we have, for ease of notation,
formulated the above result for two subsets $A$ and $B$ only, the proof
obviously goes through for any collection $A_1, \ldots, A_K \subset S$,
if the $J(A_k)$, $1 \leq k \leq K$, are pairwise disjoint. 
We then obtain that
$\{\pi^{}_{A_1} . Z_t\}_{t \geq 0}, \ldots, \{\pi^{}_{A_K}.Z_t\}_{t \geq 0}$ are
conditionally independent
Markov chains.
If, furthermore, $\pi^{}_{A_1} . Z_0, \ldots, \pi^{}_{A_K}.Z_0$ are independent,
conditional independence of the marginal processes turns into independence.
In particular, this is the case if $Z_0$ is fixed 
(since $\PP(Z_0 \in \cdot)$ is then a point
measure on some $s \in E$).
\end{remark}

As an immediate consequence, we now arrive at

\begin{theorem}
\label{thm:reco_expect}
Let $\{Z_t\}_{t \geq 0}$ be the recombination process without resampling (i.e., $b=0$), and let $Z_0$ be fixed. Then,
$\mathbb{E} (Z_t )$ satisfies the differential equation
$$
\frac{d}{dt} \mathbb{E} (Z_t ) =  \varPhi \left( \mathbb{E} (Z_t) \right)
$$
with initial value $Z_0$, and
$\varPhi$  from \eqref{eq:compact}; therefore,
\[
  \EE(Z_t) = \varphi_t(Z_0), \quad \text{for all} \; t \geq 0,
\]
with $\varphi_t$ from \eqref{eq:recosol}. 
\end{theorem}

\begin{proof}
Applying Fact \ref{fact:exp_change}, Eq.~\eqref{eq:recombinator}, 
Lemma~\ref{lem:indep_marginals}, and the linearity of the expectation,
one finds
\begin{eqnarray*}
\frac{d}{dt} \mathbb{E} (Z_t) & = & \sum_{\alpha \in L}  
\varrho^{}_{\alpha}\mathbb{E} \big ( (R_\alpha - \one) Z_t \big ) \\
& = & \sum_{\alpha\in L}  
\frac{\varrho^{}_\alpha}{N} \big [ \mathbb{E} \big ( (\pi^{}_{<\alpha} 
. Z_t) \otimes (\pi^{}_{>\alpha} . Z_t) \big ) - \EE(Z_t) \big ]\\
& = & \sum_{\alpha\in L}  
\frac{\varrho^{}_\alpha}{N} \big [ \big ( \mathbb{E} (\pi^{}_{<\alpha} 
. Z_t) \big ) \otimes \big ( \mathbb{E} ( \pi^{}_{>\alpha}. Z_t) \big) 
- \EE(Z_t) \big ] \\
& = & \sum_{\alpha\in L}  
\frac{\varrho^{}_\alpha}{N} \big [ \big ( \pi^{}_{<\alpha} . \mathbb{E} ( 
 Z_t) \big ) \otimes \big (\pi^{}_{>\alpha}. \mathbb{E} (  Z_t)  \big )
- \EE(Z_t) \big] \\
& = & \varPhi \big ( \EE(Z_t) \big ).
\end{eqnarray*}

(Clearly, the third step is the decisive one; it hinges on the independence 
of the marginals.)  
\end{proof}

\begin{remark}\label{rem:oneoff}
Note that Theorem~\ref{thm:reco_expect} does {\em not} hold for
the alternate recombination scheme \eqref{eq:rate_oneoff}. This
is because the associated resampling effect already 
destroys the validity of Lemma \ref{lem:indep_marginals},
which is essential for the proof.
We will come back to this in the next paragraph; but let us already
mention here that the deterministic solution
continues to be an excellent approximation to the expectation,
as long as recombination rates are small.
It will also be shown below that,
in practice,  the resampling effect thus introduced is
minor as long as recombination rates are small, and averages continue to be
very close to the deterministic solution.
\end{remark}

Now that the equivalence between the deterministic solution and the
expectation has been safely established at the 
level of the type frequencies, it
immediately carries over to the linkage disequilibria:

\begin{coro}
Under the assumptions  of Thm.~\ref{thm:reco_expect}, one has, for all $t \geq 0$,
\[
  \EE(T_G Z_t) = T_G \big (\varphi_t(Z_0) \big ).
\]
\end{coro}

\begin{proof}
Use the definitions of $T_G$ and $R_H$, then Remark~\ref{rem:mutual},
and finally Theorem~\ref{thm:reco_expect}. 
\end{proof}

In particular, if we define
\begin{equation}\label{eq:stoch_LDE}
 C_t \left( \langle j_1 ,\dots ,j_k \rangle\right) 
:= \left( T_{\{ \alpha <j_1 \} 
\cup \{ \alpha >j_k\}} (\hZ_t) \right) 
\left( \langle j_1,\dots ,j_k\rangle\right)
\end{equation}
(so that $C_t$ is the stochastic equivalent of $d_t$ of \eqref{eq:d_t}),
we obtain the relation $\EE(C_t\left( \langle j_1 ,\dots ,j_k \rangle\right)) = 
d_t\left( \langle j_1 ,\dots ,j_k \rangle\right)$, for all
cylinder sets $\langle j_1 ,\dots ,j_k \rangle$.

\begin{figure}[ht]
\psfrag{t}{}
\begin{center}
\subfigure[haplotype frequency, single trajectories]{\label{haplo:single}
\includegraphics[width=.48\textwidth]{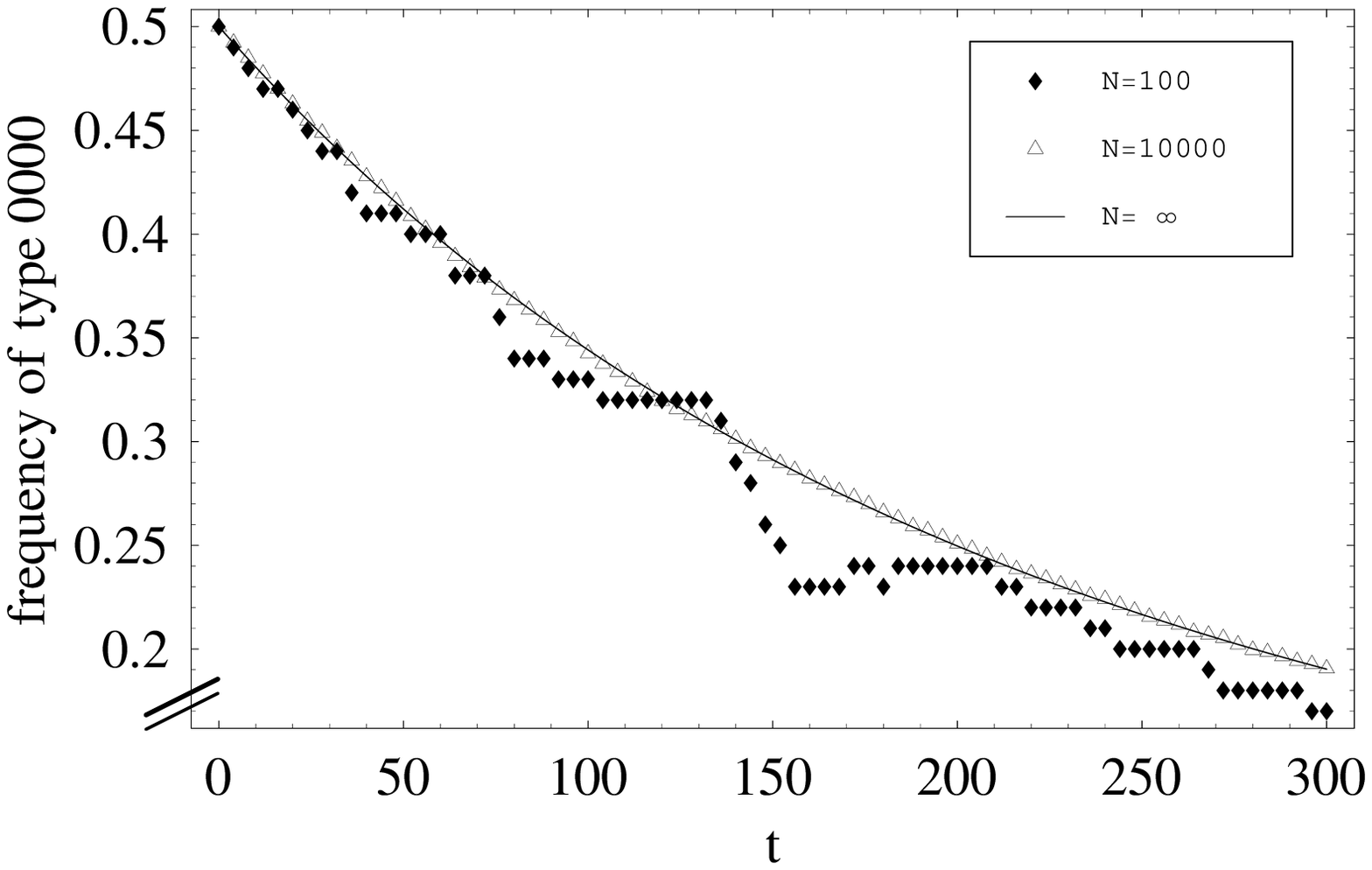}} \hfill
\subfigure[haplotype frequency, averaged over $100$ runs]{\label{haplo:av100}
\includegraphics[width=.48\textwidth]{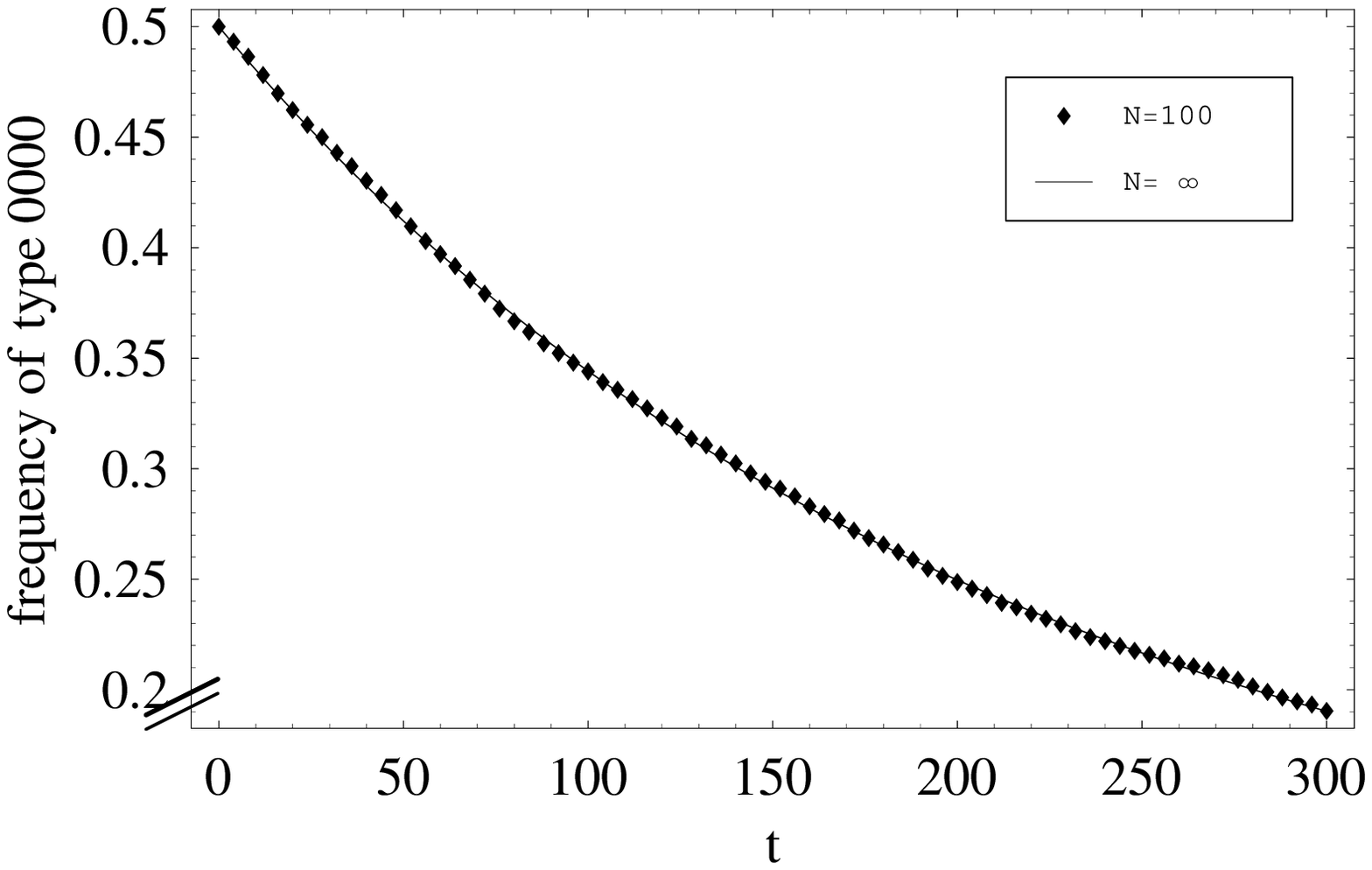}} \hfill
\subfigure[as \ref{haplo:single}, but with one offspring only]{\label{haplo:1off:1traj}
\includegraphics[width=.48\textwidth]{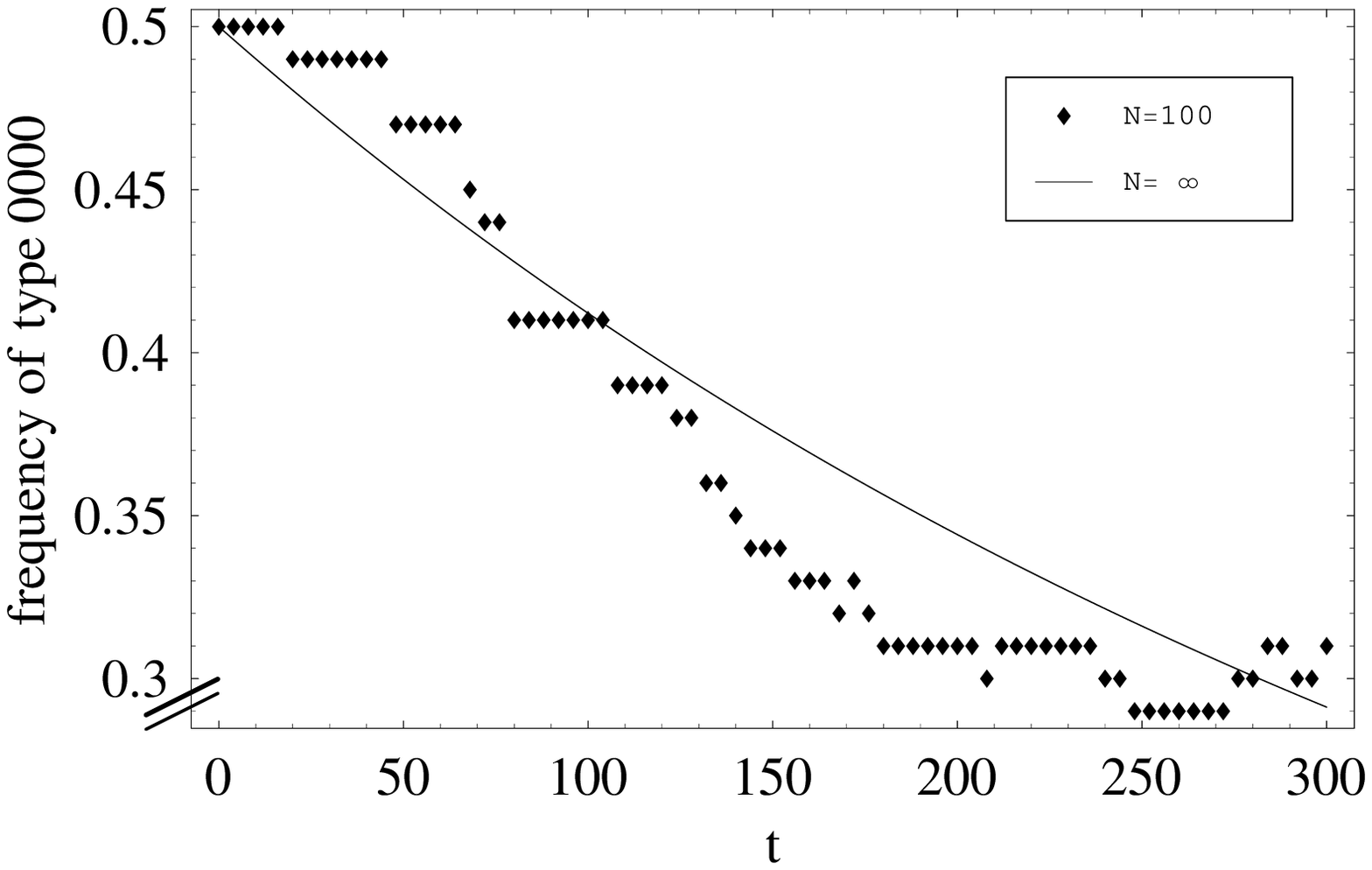}} \hfill
\subfigure[as (b), but with one offspring only]{\label{haplo:1off}
\includegraphics[width=.48\textwidth]{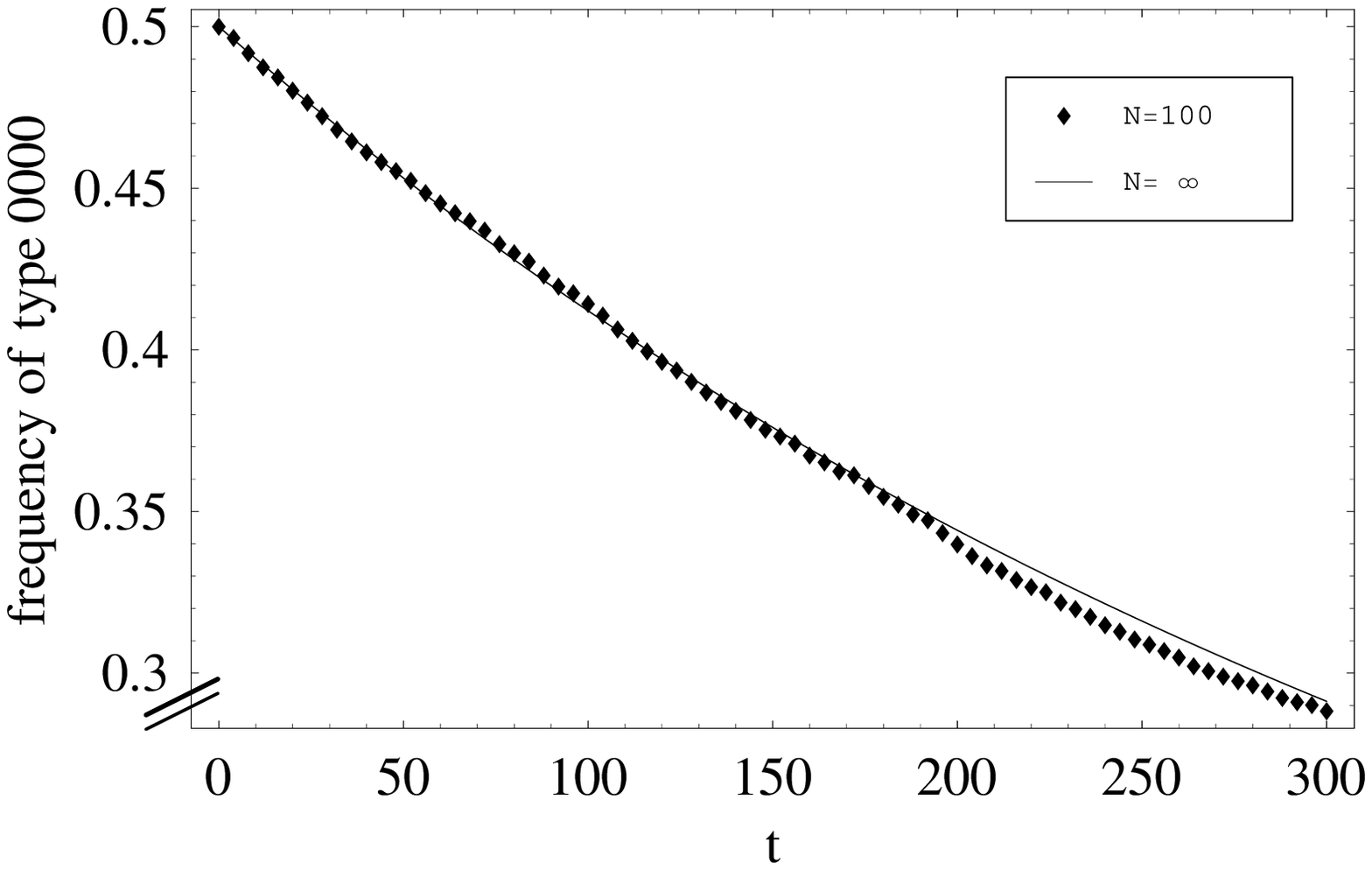}} \hfill
\subfigure[LDE, single trajectories]{\label{lde:single}
\includegraphics[width=.48\textwidth]{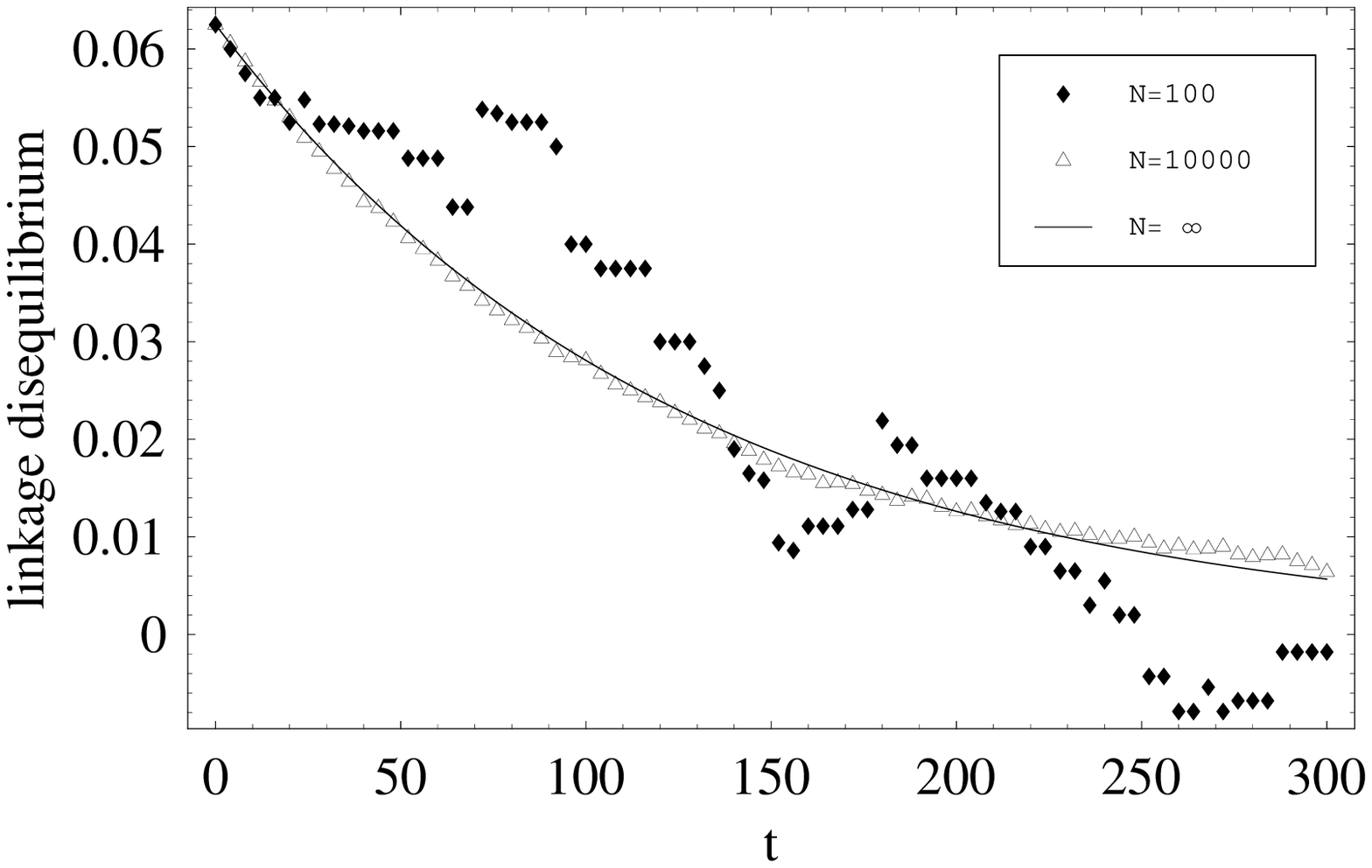}}\hfill
\subfigure[LDE, averaged over $100$ runs]{\label{lde:av100}
\includegraphics[width=.48\textwidth]{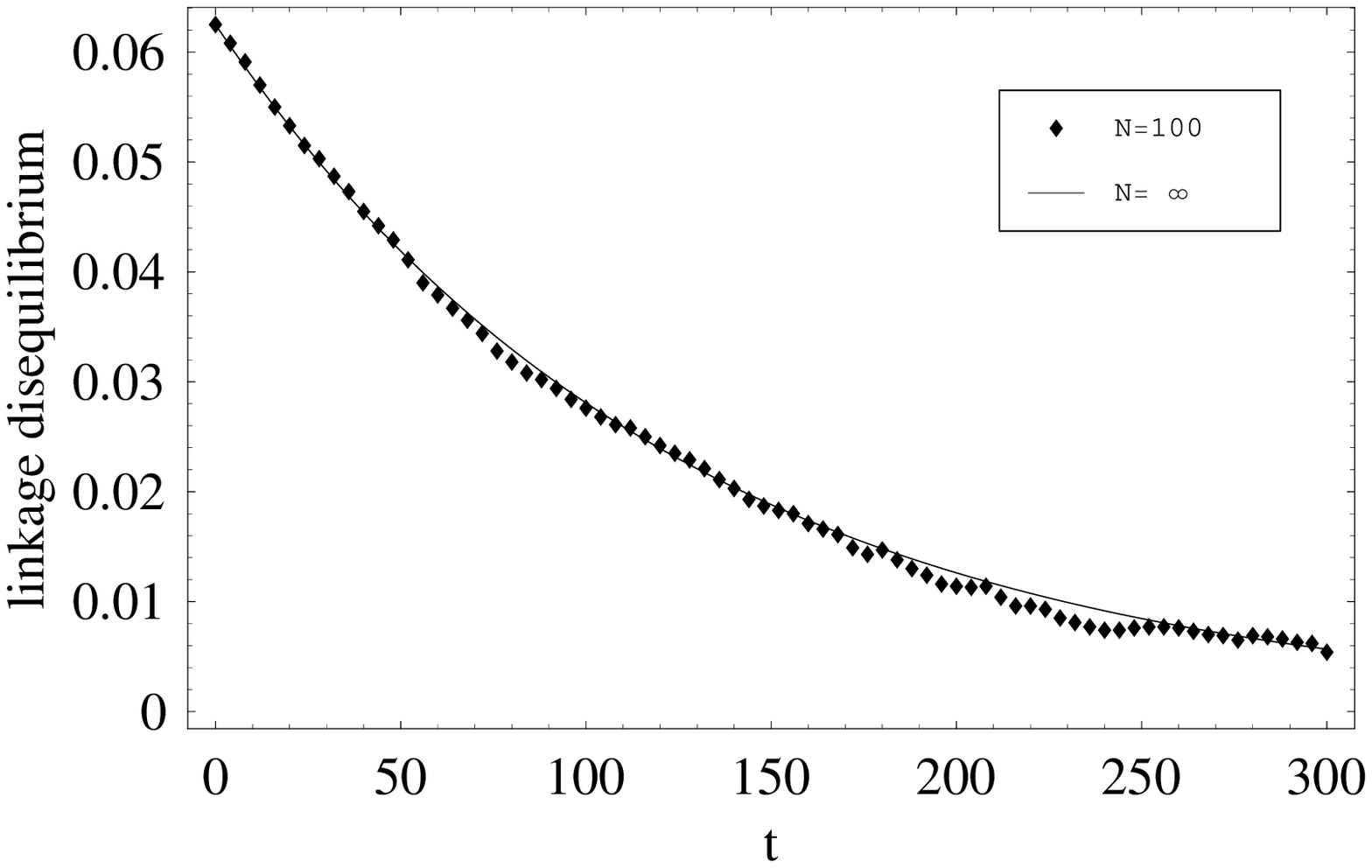}}

\caption{ {\footnotesize 
Recombination without resampling in a 4-locus 2-allele system.
Sites: $S=\{0,1,2,3\}$, types: $X=\{0,1\}^4$; recombination
rates: $\rho=0.008,
\varrho^{}_{1/2}=\varrho^{}_{3/2}=\varrho^{}_{5/2} = \rho/3$;
initial conditions:
$Z_0(0000)=Z_0(1111)= N/2$ (no other types are present).
Horizontal axis: time ($t$).
Panels \ref{haplo:single} -- \ref{haplo:1off} show realisations of $\hZ_t(0000)$ for various population sizes, as
single trajectories \ref{haplo:single}, or means over 100 \ref{haplo:av100} 
realisations; also shown is
$p_t(0000)=(\varphi_t(\hZ_0))(0000)$ (the deterministic solution or
$N \to \infty$ limit), which equals $\EE(\hZ_t)$ by 
Thm.~\ref{thm:reco_expect} (and is partly hidden between the diamonds in 
\ref{haplo:av100} and \ref{haplo:1off}). 
Clearly, the sample mean is well described by
the expectation, where averaging is faster in larger populations.
Panel \ref{haplo:1off:1traj} shows a single trajectory, and
panel \ref{haplo:1off} an average over 100 realisations, for a simulation
according to Eqs.~\eqref{eq:trans1_oneoff}--\eqref{eq:rate_oneoff}, 
in which only one offspring individual
is produced, rather than two as in Eq.~\eqref{eq:rate_reco}, the
assumption underlying all other simulations.
Obviously, this only  introduces a minor resampling effect and corresponding
systematic deviation from the deterministic limit (but note that,
relative to the usual resampling scheme, 
the dynamics is slowed down by a factor of $1/2$). 
Panels \ref{lde:single} and \ref{lde:av100} show  $C_t(\langle 0,1,2,3\rangle)$ 
(of Eq.~\eqref{eq:stoch_LDE}), for type 0000, and  various population sizes,
as single trajectories \ref{lde:single}, or means \ref{lde:av100} over 100 realisations,
and compares them with the deterministic quantity (or $N \to \infty$ limit)
$d_t(\langle 0,1,2,3\rangle)$. 
}}
\label{fig:reco_only}
\end{center}
\end{figure}

Fig.~\ref{fig:reco_only} illustrates the findings of this paragraph
for a four-locus two-allele system with recombination and no resampling.
We have chosen $\rho=0.008$ and $\rho_{\alpha}=\rho/3$ for 
$\alpha \in \{\frac{1}{2},\frac{3}{2},\frac{5}{2} \}=L$, which corresponds
to four sites spaced evenly across a stretch of $8 \cdot 10^5$ bp,
at a per-nucleotide recombination rate of $10^{-8}$.  
The Figure shows haplotype frequencies and
highest-order LDEs, both as single trajectories and averages over many
realisations, as compared to the  solution of the deterministic recombination equation. For very small populations ($N=100$), single trajectories fluctuate markedly,
but averages over 100 realisations are already indistinguishable from
the deterministic solution---in line with the property of the expectation just established.
For larger populations, the stochasticity is already greatly reduced in single
trajectories---but this is the topic of the next paragraph.

The Figure also contains results from the alternate resampling
scheme with one offspring only, \eqref{eq:trans1_oneoff} --
\eqref{eq:rate_oneoff}.
Although Theorem~\ref{thm:reco_expect} is not exactly valid in this
case,  the resampling effect thus introduced is obviously
minor as long as recombination rates are small, and averages continue to be
very close to the deterministic solution.

\makeatletter
\renewcommand{\p@subfigure}{}
\subfigcapskip -3ex
\subfigcapmargin +2ex
\makeatother

\subsection{Recombination and resampling: the infinite population limit}
Let us now include resampling, at rate $b/2>0$, and consider the
stochastic process $\{Z_t^{(N)}\}_{t \geq 0}$ defined by both 
\eqref{eq:rate_resampling}
and \eqref{eq:rate_reco}, where we add the upper index here to indicate
the dependence on $N$. Now, $\varPhi$ and $\EE$ no longer commute. 
The processes $\{\pi^{}_{<\alpha}.Z_t\}_{t \geq 0}$ 
and $\{\pi^{}_{>\alpha} . Z_t\}_{t \geq 0}$ are still individually Markov,
but their
resampling events are  coupled (replacement of $y^{}_{<\alpha}$ by $x^{}_{<\alpha}$
is always tied to replacement of $y^{}_{>\alpha}$ by $x^{}_{>\alpha}$).
Hence the marginal processes fail to be independent, so that
no equivalent of Lemma
\ref{lem:indep_marginals} holds.

Let us, therefore, change focus and consider the normalised version 
$\{\hZ_t^{(N)}\}_{t \geq 0} = \{Z_t^{(N)}\}_{t \geq 0}/N$.
It seems to be general folklore in population genetics
that, in the limit $N\to \infty$, the 
relative frequencies  of the Moran or Wright-Fisher model
cease to fluctuate and are then given by the solution of the 
corresponding deterministic equation. This is implied no matter 
which evolutionary
forces (like mutation, selection, or recombination) are included
into the model; in our case, $\{\hZ_t^{(N)}\}_{t \geq 0}$ should
be described by the differential equation \eqref{eq:ode} as $N\to \infty$.
However, the limit theorem behind this 
is usually not made explicit, and, in fact, does not seem
to be   well known in the population genetics literature. 
Indeed, it is given by a very general law of large numbers due to 
Ethier and Kurtz
\cite[Thm.~11.2.1]{EtKu86}, which provides 
the formal justification for a very large 
class of  models in biology that are stochastic at the microscopic
scale but are adequately described deterministically if the population
size is large; they range from biochemical reaction 
kinetics to population dynamics and population genetics. For our special 
case, it reads as follows. 
\begin{prop}\label{prop:lln}
Consider the family of processes
$\{\hZ_t^{(N)}\}_{t \geq 0}=\frac{1}{N} \{Z_t^{(N)}\}_{t\geq 0}$,
$N=1,2,\ldots$,  where
$\{Z_t^{(N)}\}_{t\geq 0}$ is defined by 
\eqref{eq:rate_resampling} and \eqref{eq:rate_reco}.
Assume that the initial states are chosen so that
$\lim_{N\to\infty} \hZ^{(N)}_0 =p_0$. Then, 
for every given $t\geq 0$, one has
\be{eq:LLN}
\lim_{N\to\infty} \sup_{s\leq t} | \hZ^{(N)}_s - p_s | =0 
\ee
with probability $1$, where $p_s := \varphi_s(p_0)$ is the
solution of the deterministic recombination equation \eqref{eq:ode}.
\end{prop}
\begin{proof}
To apply Thm.~11.2.1 of \cite{EtKu86}, we need a linear scaling
of the transition rates with $N$.
More precisely, we must show that the transition rates 
$Q^{(N)}(z,z+v)$ of the 
process $\{Z^{(N)}_t\}_{t \geq 0}$ are, for all $z, z+v \in E$, 
of the form 
\begin{equation}\label{eq:scaling}
Q^{(N)}(z,z+v) = Nq_v(z/N), \quad \text{ for all } N,
\end{equation}
with non-negative functions $q_v$  
defined on a subset of $\RR^{|X|}_{\geq 0}$. 
Setting 
\begin{eqnarray}\label{eq:qv_s}
 q_v^{(s)}(p)& := & \frac{b}{2}  \sum_{\substack{(x,y) \in X \times X:\\
                  s(x,y)=v}}  p(x) p(y), \\
 \label{eq:qv_r}
 q_v^{(r)}(p) & := & \frac{1}{2} \sum_{\substack{(x,y) \in X \times X, \alpha \in L:\\
                  r(x,y,\alpha)=v}}  \varrho^{}_{\alpha} p(x) p(y), 
\end{eqnarray}
together with $q_v := q_v^{(s)}+q_v^{(r)}$, and recalling that 
\[
  Q^{(N)}(z,z+v)=Q^{(N,s)}(z,z+v)+Q^{(N,r)}(z,z+v)
\] (from \eqref{eq:Q_s} and
\eqref{eq:Q_r} -- now with notational dependence on $N$), 
it is obvious that \eqref{eq:scaling} is indeed satisfied.
(Observe that this just reflects the relation
$z(x)z(y)/N= N \frac{z(x)}{N} \frac{z(y)}{N}$.)
The normalised process $\{\hZ^{(N)}\}_{t \geq 0}$ on $E/N$
has the corresponding transitions 
$z/N \to z/N + v/N$, again at rates $Nq_v(z/N)$. Thus, 
the collection of processes  $\{\hZ^{(N)}\}_{t \geq 0}$ 
constitutes what is known as a {\em density-dependent
family corresponding to the $q_v$} \cite[p.~455]{EtKu86}.

Now, for such a density-dependent family, Thm.~11.2.1 of
\cite{EtKu86} implies \eqref{eq:LLN} 
if $p_t$ solves $\dot p_t = f(p_t)$, with initial value $p_0$,
and $f(p) := \sum_v v q_v(p)$. Proceeding as in \eqref{eq:F_def} and
\eqref{eq:F=Phi}, we obtain
$f(p) =  \varPhi(p)$ in analogy with 
Fact~\ref{fact:F=Phi}. 
\end{proof}

Note that the convergence in \eqref{eq:LLN} 
applies for any given $t$,
but need not hold  for $t \to \infty$; we shall come back to this
point in the Discussion.
Note also that
the convergence carries over to linkage disequilibria, 
since $T_G(\hZ_t)$ converges to $T_G(p_t)$ (in the above sense).

The question that remains is whether this limit result bears any
relevance to real populations, which are always finite.
How large must
$N$ be for the infinite-population limit to yield a reasonable approximation?
 
\begin{figure}[ht]
\psfrag{t}{}
\begin{center}
\subfigure[haplotype frequency, single trajectories]{\label{res:haplo:single}
\includegraphics[width=.48\textwidth]{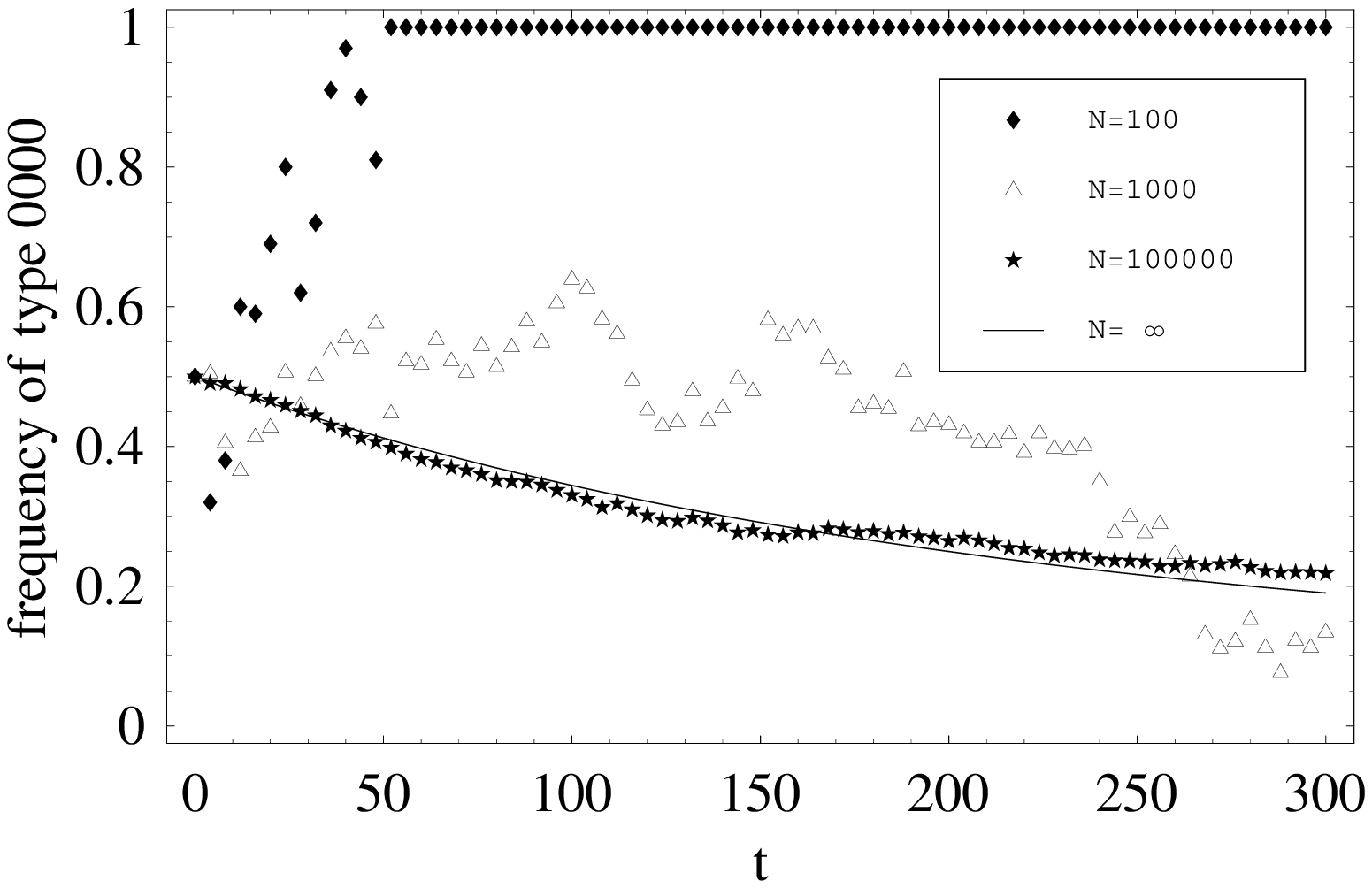}} \hfill
\subfigure[haplotype frequency, averaged over $100$ runs]{\label{res:haplo:av100}
\includegraphics[width=.48\textwidth]{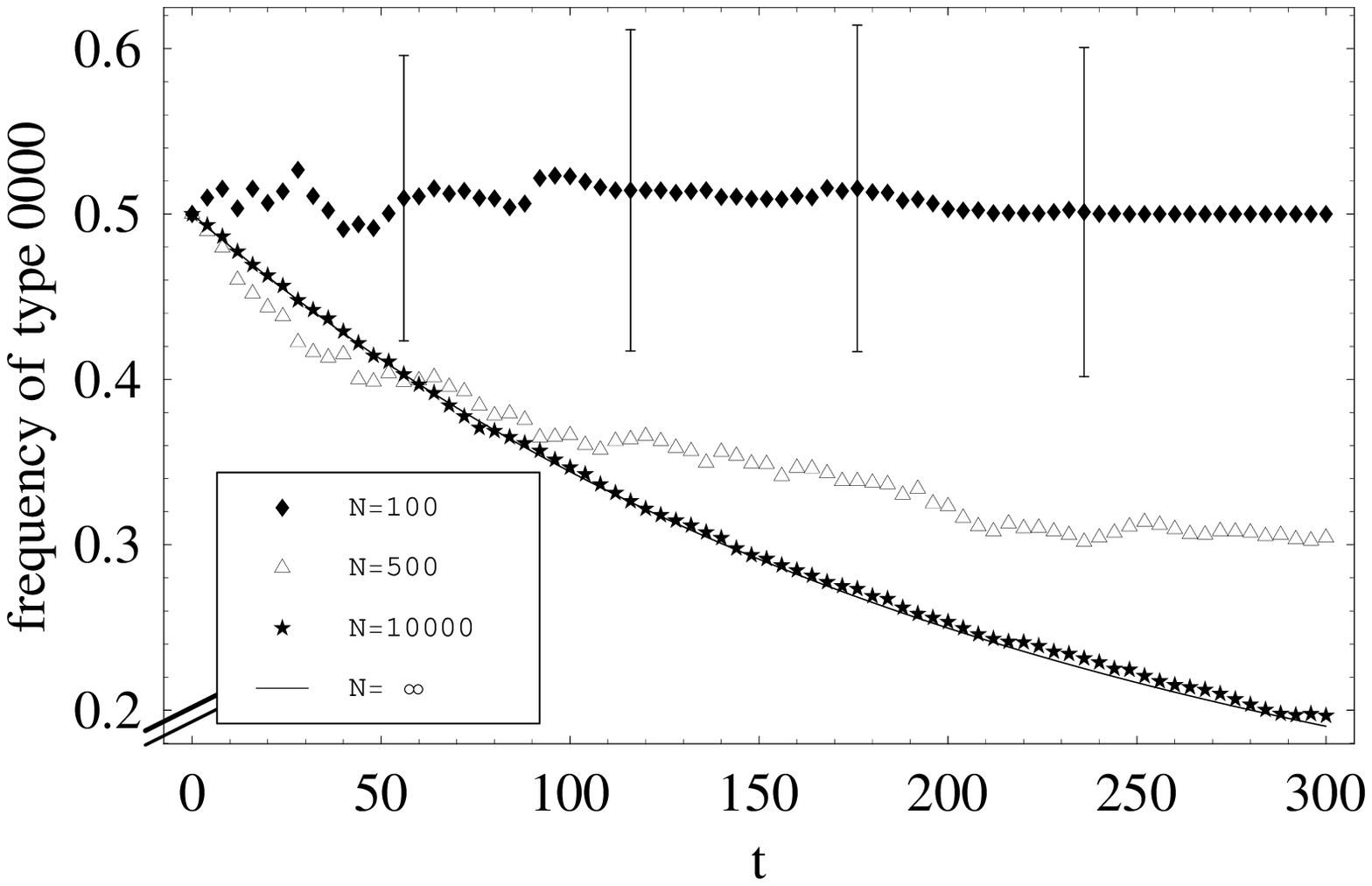}} \hfill
\subfigure[LDE, single trajectories]{\label{res:lde:single}
\includegraphics[width=.48\textwidth]{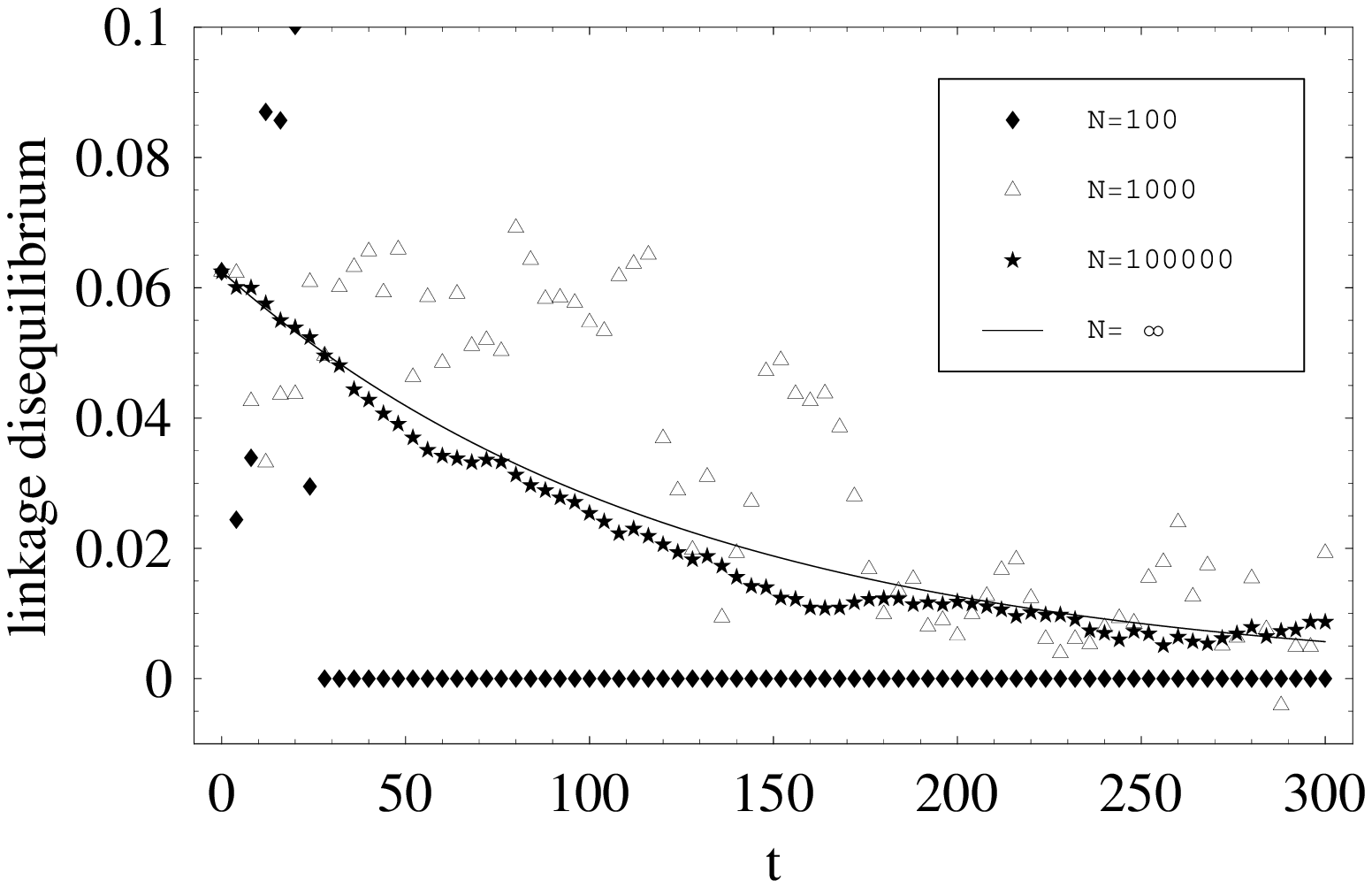}} \hfill
\subfigure[LDE, averaged over $100$ runs]{\label{res:lde:av100}
\includegraphics[width=.48\textwidth]{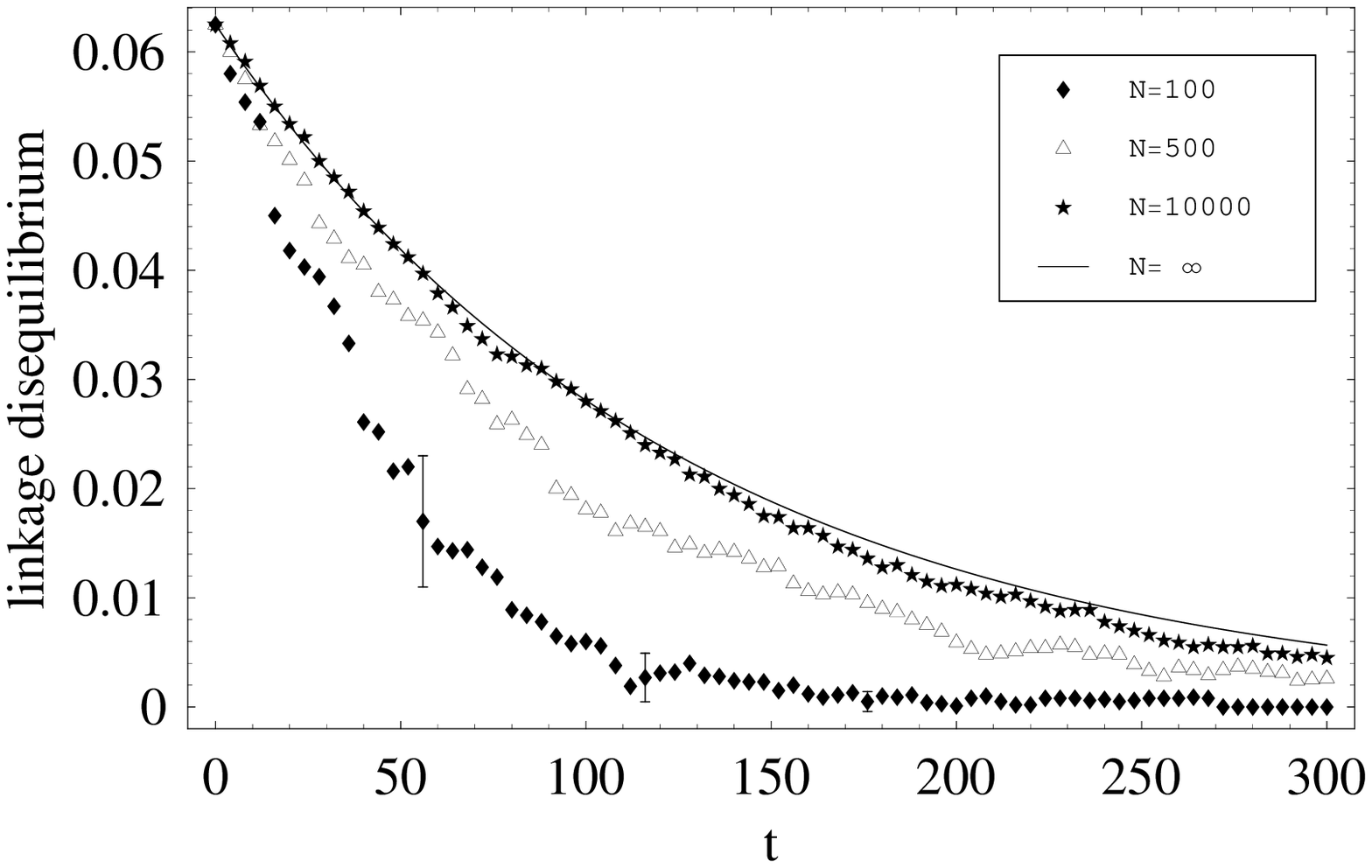}}
\caption{{\small Recombination with resampling in the 4-locus 2-allele system.
Sites, types,  recombination rates, and initial 
conditions as in Fig.~\ref{fig:reco_only}. Resampling rate: $b=2$.
 Horizontal axis: time ($t$).
Panels \ref{res:haplo:single} and \ref{res:haplo:av100}  show realisations of $\hZ_t(0000)$ for various population sizes,
as individual trajectories \ref{res:haplo:single} or averages over 100 realisations
\ref{res:haplo:av100}; also shown is 
$p_t(0000)=(\varphi_t(\hZ_0))(0000)$ (the corresponding deterministic solution, or $N \to
\infty$ limit), partly hidden between the $N=10000$ stars in
\ref{res:haplo:av100}. 
Single
trajectories \ref{res:haplo:single} approach the infinite population limit already  for
moderate population
sizes ($N=10^5$). In contrast, averages \ref{res:haplo:av100}
deviate strongly from the deterministic solution if the population
is small (error bars on the $N=100$ curve 
correspond to 95 \% confidence intervals), but are hard to distinguish from the
infinite population limit for $N=10000$. Analogous results hold
true for single trajectories \ref{res:lde:single}, and averages over 100 realisations
\ref{res:lde:av100}, of the highest-order LDE, $C_t(\langle 0,1,2,3\rangle)$,
evaluated for type 0000; the $N =
\infty$ line is $d_t$. } 
}
\label{fig:reco_res}
\end{center}
\end{figure}

We have investigated this by means of simulations of our four-locus
two-allele system
(Fig.~\ref{fig:reco_res}). As is to be expected, single realisations, as well as averages, 
approach the deterministic limit with increasing $N$;
and for 
$N=10^5$, the stochastic process is already very close to the
limit. This is observed at the level of type frequencies, as well as linkage 
disequilibria. The situation is very similar to that
without resampling
(Fig.~\ref{fig:reco_only}, (a) and (e)), except that somewhat larger
population sizes are required due to the increased stochasticity
induced by  resampling. In contrast, 
averages over {\em small} populations are not expected to, and in
fact do not,  get close to
the deterministic solution.


\section{Discussion}
The main purpose of this article has been to clarify some relationships
between the Moran model with recombination and the deterministic
recombination model (both with single cross\-overs). To separate
the effects of recombination from those of resampling, we
formulated the Moran model in its `decoupled' version, with
independent recombination and resampling events; this approach
is also taken in \cite{PHW06}, for example. 
The coupling of recombination to resampling (which happens to be
biological reality) is thus neglected. Put differently, our model
describes correctly the resampling effects due to reproduction events
that do not involve recombination, but neglects those resampling events
that occur in the course of reproduction associated
with recombination. But recombination is rare (relative to
reproduction), at least at the molecular level aimed at by
the single-crossover model; and the bias introduced
by this simplification is accordingly small, 
as also illustrated by our simulations of the
alternative recombination scheme \eqref{eq:trans1_oneoff} --
\eqref{eq:rate_oneoff}.

As the main result of this article, we have shown:
\begin{enumerate}
\item For recombination without resampling, the expected type frequencies
are given by the deterministic dynamics, for arbitrary (even small)
population sizes. Although this is not a biologically realistic situation,
it yields  insight into the Moran model with recombination, and
establishes a relationship between finite
and infinite populations that is somewhat unexpected in view
of the inherent nonlinearity of recombination. 
The key observation that led to this result rests upon the fact that
the marginal processes (i.e., the type frequencies at the sites
before resp.\ after a given link) are independent Markov chains.
Of course, this is related to the fact that, in the absence of resampling,
the genetic material is completely conserved (just rearranged);
in particular, types cannot go to fixation. 

\item 
The combined model {\em with} resampling deviates 
(in expectation) from the deterministic dynamics; in view of
the findings above,  these deviations are solely
due to resampling, rather than  recombination.
But the infinite population limit continues to apply, 
and is, again, given by the deterministic recombination equation.
We have investigated its
range of validity here for one representative scenario 
(four biallelic loci,  a time horizon
of $T=300$,  an expected number of 
$T \varrho^{}_{\alpha}=300 \cdot 0.008/3 = 0.8$ recombination
events per link and individual, and $T  b/2=300$ resampling 
events per individual); then, 
a population of moderate size ($N=10^5$) is
already close to the deterministic limit. It should be noted, however,
that this result is expected to further vary with:

\begin{enumerate}
 \item
 The number of sites, or of alleles per site (or, more generally,
 the size of the type space):  
 The deterministic limit can only be  a good
 approximation if {\em all types} are well-populated; for a
 larger type space, the required population size will be larger.
 But situations with four loci (or variable nucleotide sites) already 
 cover many interesting biological situations; and typing them
 (and determining the corresponding four-way LDEs) is, after all,
 a veritable  task that yields a great deal of information.
It remains to be investigated, however,  how the
quality of the approximation 
changes when there are more than two alleles per site.

\item
The time horizon: The law of
large numbers  \eqref{eq:LLN} holds for every given, finite time horizon,
but need not carry over to 
$t \to \infty$. Indeed, if resampling is present, the population size
required to get close to the deterministic solution
is expected to grow over all bounds with increasing $t$. 
This is because, for every finite $N$, the Moran model with resampling and
recombination is an {\em absorbing} Markov chain, which leads to
fixation (i.e., a homogeneous population of uniform type) in
finite time with probability one (for the special case of just two
types without recombination, the expected time 
is of order
$N$, if the initial frequencies are both $1/2$ \cite[p.~93]{Ewen04}).
In sharp contrast, the
deterministic system never loses any type, and the stationary state,
the complete product measure \eqref{eq:stat_state},
is, in a sense, even the most variable state accessible to
the system. For increasing $N$, 
finite populations stay close to the deterministic limit for an increasing
length of time (see Fig.~\ref{fig:reco_res}). Indeed, our main
interest here is this time horizon during which substantial changes in LDE
occur, and this is described by the deterministic model;
in contrast,  the deterministic
solution does not provide a meaningful picture for the equilibrium
situation.
The case would be very different if mutation were included into
the model, since this would turn the absorbing Markov chain into
an ergodic one, whose stationary distribution allows a meaningful
comparison with the deterministic dynamics even for $t \to \infty$.
\end{enumerate}
\end{enumerate}

Let us finally discuss implications for the corresponding model
in discrete time, that is, the Wright-Fisher model with recombination.
Again, we may consider
\begin{itemize}
\item a model without resampling (i.e., the only
events  are single crossovers between pairs of
individuals, with pairs of offspring replacing pairs of parents): 
Although one might assume this model to behave like
the corresponding discrete-time dynamical system for finite $N$,
{\em in expectation},
we indeed stipulate that this is not the case. The reason is that,
in discrete time, the marginal processes (related to non-overlapping
stretches) cease to be independent. This is because such independence
would imply the possibility of two or more crossovers in one
generation -- in contrast to the single-crossover assumption.
For the same reason, 
the deterministic single-crossover model in discrete time is
much more difficult to solve than in continuous time, see 
the discussion in \cite{BaBa03}.
\item recombination combined with resampling: It is strongly
expected that,
for $N \to \infty$, there is again a law of large numbers,
which makes the relative frequencies converge to the corresponding
deterministic dynamics, in analogy with Prop.~\ref{prop:lln}.
However, no explicit equivalent of the underlying Theorem 11.2.1 of
\cite{EtKu86} is known to the authors.
\end{itemize}

\subsection*{Acknowledgments}
It is our pleasure to thank Richard Hudson for a decisive hint
on the structure of stochastic recombination,
and the ICMS workshop on Mathematical Genetics (Edinburgh 2006) 
for providing ample opportunity to 
discuss the results.
We are grateful to Michael Baake, Matthias Birkner, and
Thiemo Hustedt for critically reading the manuscript.
This work was supported by the Dutch-German Bilateral Research
Group on Random Spatial Models in Physics and Biology (DFG-FOR 498).





\newcommand{\noopsort}[1]{} \newcommand{\printfirst}[2]{#1}
  \newcommand{\singleletter}[1]{#1} \newcommand{\switchargs}[2]{#2#1}

\end{document}